\documentclass[twocolumn,amsmath,amssymb,superscriptaddress,aps,pra]{revtex4-2}
\pdfoutput=1

\usepackage[utf8]{inputenc}
\usepackage{graphicx}
\usepackage{hyperref}
\usepackage{amsmath}
\usepackage{amssymb}
\usepackage{xcolor}
\usepackage{amsthm}
\usepackage{cleveref}
\usepackage{tikz}
\usepackage{quantikz}
\usepackage{multirow}
\usepackage{array}

\begin{document}

\title{To reset, or not to reset -- that is the question}
\author{Gy\"{o}rgy P. Geh\'{e}r} 
\affiliation{Riverlane, Cambridge, CB2 3BZ, UK}
\author{Marcin Jastrzebski}
\affiliation{Riverlane, Cambridge, CB2 3BZ, UK}
\affiliation{Department of Physics and Astronomy, University College London, London WC1E 6BT, UK}
\author{Earl T. Campbell}
\affiliation{Riverlane, Cambridge, CB2 3BZ, UK}
\affiliation{School of Mathematical and Physical Sciences, University of Sheffield, Sheffield S3 7RH, UK}
\author{Ophelia Crawford}
\affiliation{Riverlane, Cambridge, CB2 3BZ, UK}


\begin{abstract}
Whether to reset qubits, or not, during quantum error correction experiments is a question of both foundational and practical importance for quantum computing.  Text-book quantum error correction demands that qubits are reset after measurement.   However, fast qubit reset has proven challenging to execute at high fidelity.  Consequently, many cutting-edge quantum error correction experiments are opting for the no-reset approach, where physical reset is not performed.  It has recently been postulated that no-reset is functionally equivalent to reset procedures, as well as being faster and easier.  For memory experiments, we confirm numerically that resetting provides no benefit.  On the other hand, we identify a remarkable difference during logical operations. We find that unconditionally resetting qubits can reduce the duration of fault-tolerant logical operation by up to a factor of two as the number of measurement errors that can be tolerated is doubled. We support this with numerical simulations. However, our simulations also reveal that the no-reset performance is superior if the reset duration and infidelity exceed given thresholds. For example, with the noise model we considered, we find the no-reset performance to be superior when the reset duration is greater than approximately $100$ ns and the physical error probability is greater than approximately $10^{-2.5} \approx 0.003$. Lastly, we introduce two novel syndrome extraction circuits that can reduce the time overhead of no-reset approaches. Our findings provide guidance on how experimentalists should design future experiments.
\end{abstract}

\maketitle


Quantum error correction (QEC) is a vital tool for unlocking quantum applications far beyond the capabilities of classical computers. QEC works by periodically measuring a set of multi-qubit Pauli operators, called stabilisers, providing evidence of errors that are then corrected. Measuring a stabiliser commonly involves a so-called syndrome extraction circuit that uses auxiliary qubits to assist with measurements on the data qubits that store the logical information. Many syndrome extraction circuits start by resetting the auxiliary qubits to the $|0\rangle$ state and end with single-qubit readout of the auxiliary qubits in the $Z$ basis, thereby collapsing to either the $|0\rangle$ or $|1\rangle$ state~\cite[e.g.,][Sec. 4.4]{NielsenChuang}. As these circuits are repeatedly used, the collapsed states on the auxiliary qubits need to be reset back to $|0\rangle$. Resetting can be achieved in various ways that seem, at first appearance, to be equivalent. However, in this paper, we reveal fundamental differences in the fault-tolerant properties of reset schemes previously thought to be similar. 

Let us survey the three broad classes of reset strategy. The first class is \textit{unconditional reset}, where a non-unitary operation maps all qubit states into the $|0\rangle$ state. The simplest option to achieve this is to wait until the qubits decay into their ground states \cite{DiVincenzo}. However, this is too slow for mid-circuit reset.  By engineering a suitable interaction between a qubit and its environment, energy can be dissipated faster and with higher fidelity \cite{fast-and-unconditional-reset, rapid-unconditional-reset}. Unconditional reset is the conventional text-book approach and in the QEC literature is usually just called reset. The second class is \textit{conditional reset}~\cite{conditional-reset}, where a conditional bit-flip gate is applied on the auxiliary qubit if the previous measurement outcome was $1$, otherwise no gate is applied.  The duration of conditional reset depends on both the bit-flip gate duration and the classical electronics latency, which increase the execution time of the circuit. However, alternatively, we may simply track the effect of the conditional bit-flip in software, which is exactly what is involved in the third class, \textit{no reset} \cite{Delft-13, Delft-planar-docde, ETH-planar-code}.  Recent years \cite{campbell2024series} have seen QEC demonstrated with various qubit types \cite{quera-paper, quantinuum-paper,832-color-code-demonstration}. Focusing on superconducting qubits, several QEC experiments used the no-reset scheme \cite{Delft-13, ETH-planar-code}, while others implemented unconditional reset \cite{miao_overcoming_2022,acharya_suppressing_2023}, making it timely to investigate the differences between these approaches.  

\begin{figure}[t]
    \includegraphics{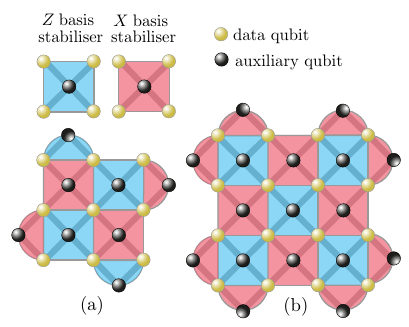}
    \caption{Stabilisers for the planar surface code for (a) distance-$3$ quantum memory and (b) width-$4$ stability experiments.  A range of distances, widths and numbers of rounds are used in our simulations.}
    \label{fig:planar_code}
\end{figure}

It has been previously claimed~\cite[Sec. S2]{miao_overcoming_2022} that choosing the no-reset scheme ``\textit{has an insignificant impact on code performance}", but this has not been fully investigated.  When using no reset or conditional reset, misclassification of measurements has two consequences:  the wrong readout outcome is recorded, \textit{and} an erroneous bit-flip (in software or on hardware) is applied to a qubit. This shows that a single failure event effectively causes a correlated pair of errors.  The central question is whether this correlated error significantly decreases the number of error events needed for an undetectable logical failure. On the one hand, we will confirm the claim from \cite{{miao_overcoming_2022}} that for quantum memory there is no significant impact on QEC performance. On the other hand, we find dramatic consequences when we perform a logical operation. As we scale quantum computers, we enter an era where it becomes possible to demonstrate small examples of fault-tolerant quantum computation (FTQC), for instance, by performing a joint Pauli measurement via lattice surgery \cite{Horsman_2012, GoSc, ChamCamtwistfree, ChamCamtwistbased,Bombin2021, geher_error-corrected_2023}. During lattice surgery, classification errors on physical qubits alone can cause an overall misclassification of the logical Pauli measurement. When classical feedback is performed conditionally on this logical Pauli measurement, logical measurement misclassification transforms into a logical qubit error. In this context, we show that in the no-reset (and conditional-reset) schemes, these logical failure mechanisms (not present in quantum memory) may be formed by half as many errors as with unconditional reset.  Consequently, to tolerate the same number of errors, no-reset schemes require doubling the duration of lattice surgery; such slow-down is clearly undesirable.

\begin{figure}
    \centering  \includegraphics[width=0.98\linewidth]{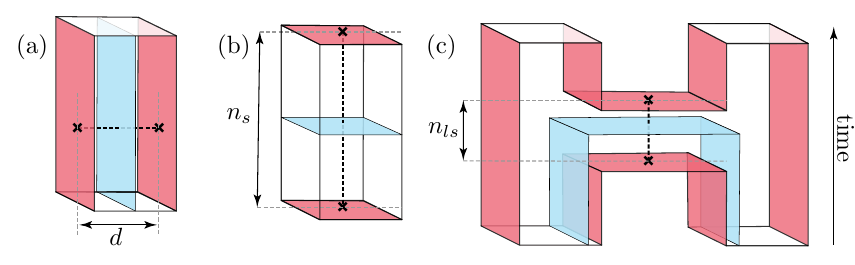}
    \caption{Topological space-time perspectives on logical failure mechanisms in QEC experiments.  (a) A quantum memory experiment and a distance $d$ undetectable logical error due to $d$ physical qubit errors.  (b) A stability experiment over $n_s$ rounds and an undetectable logical error due to $n_s$ measurement classification errors when using unconditional reset or $\lceil n_s/2 \rceil$ measurement classification errors when using no-reset. (c) A lattice surgery experiment measuring a logical $X \otimes X$ Pauli operator with the logical error shown similar to that observed during stability experiments.  In all diagrams, strings of physical qubit errors or measurement classification errors can terminate at pink boundaries.  A logical failure occurs whenever these error strings pass through the blue surfaces an odd number of times.}
    \label{fig:3D_graph_diagrams}
\end{figure}

Tolerating half as many errors will impact QEC performance, though the exact effect depends on specific details, including the noise model. We complement our aforementioned analytical insights with numerical results for a range of superconducting-inspired circuit-level Pauli noise models. Our simulations use the stability experiment~\cite{Gidney-stability} (see also \Cref{fig:planar_code,fig:3D_graph_diagrams}) as a proxy for lattice surgery and other experiments where similar logical failures occur. Our numerical results are consistent with a $2 \times$ advantage of unconditional resets in the limit of fast, low-error resets. In particular, with a physical error rate of $10^{-3}$, which can be expected in the relatively near future, we observe substantial advantage with instantaneous reset, some advantage with fast reset and no advantage with slow reset. However, for nearer-term experimental demonstrations of FTQC, when the physical error rate is higher, we observe the no-reset scheme to be, in general, superior, especially with slow resets.  We study this transition in detail, considering performance both in terms of numbers of QEC rounds and logical operation duration. 

Finally, we propose and numerically compare two alternative syndrome extraction circuits for the planar code that do not use unconditional resets but substantially decrease the aforementioned incurred time overhead. One of these novel syndrome extraction circuits uses an additional two-qubit gate per measurement round, transforming the classification error into a mixture of two more benign errors. This circuit gets close to recovering the performance of unconditional reset using the no-reset scheme. The other novel syndrome extraction circuit we propose uses two auxiliary qubits per stabiliser, effectively squeezing two measurement rounds into one. As a result, this second circuit's QEC performance is substantially better in the examined noise regime than previous circuits, although it requires $\approx50\%$ more qubits. Based on these results, we discuss when it is beneficial to use unconditional resets on superconducting hardware and when to use one of our alternative circuits.

The structure of our paper is as follows. In \Cref{sec:detectors_no_reset}, we describe how the decoding problem changes when we switch from the unconditional-reset to the conditional-reset or no-reset scheme. In \Cref{sec:qec_experiments}, we first describe a Pauli noise model, inspired by currently available realistic superconducting hardware, that we use throughout (\Cref{subsec:device}). Then, we present numerical results that verify that the differences between the three reset schemes for quantum memory with the rotated planar code (\Cref{subsec:qmem}) are small. After that, we describe the stability experiment and why it is a useful proxy for estimating the time cost of FTQC. Furthermore, we demonstrate through numerical simulations with the stability experiment that FTQC incurs a substantial time overhead in the no-reset (and conditional-reset) scheme, provided resetting is fast enough and is of high enough fidelity (\Cref{subsec:stability}). In \Cref{sec:spread,sec:swappy}, we present two alternative syndrome extraction circuits in the no-reset scheme that overcome this time overhead problem. Even though we present circuits only for the planar code, they can also be applied straightforwardly to more general QEC codes. With \Cref{sec:conclusion}, we conclude the paper by discussing the obtained results and possible future research directions.

\section{The structure of detectors in the three reset schemes}\label{sec:detectors_no_reset}
\begin{figure*}[t!]
    \centering
\includegraphics[width=0.98\textwidth]{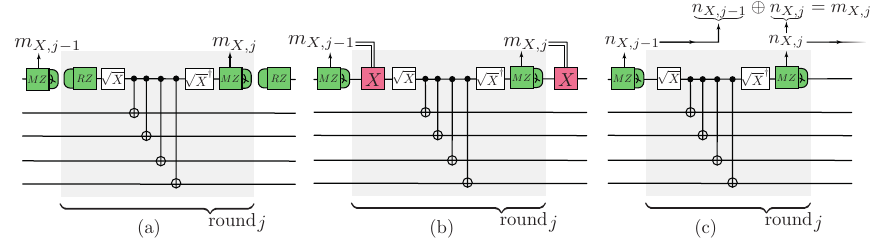}
    \caption{Standard syndrome extraction circuits for an $XXXX$ stabiliser in the three reset schemes. The circuits use $\sqrt{X}$, $\sqrt{X}^\dagger$ one-qubit gates, CX two-qubit gate, and $Z$-basis reset and measurement. (a) Standard text-book circuit using the unconditional-reset scheme. (b) The conditional-reset version of (a), where the classically-controlled bit-flip gate (pink) is applied immediately after measurement. (c) The no-reset version of (a) where the conditional bit-flip's effect from (b) is tracked in software; cf.~\Cref{eq:m_expressed_with_ns}. We note that, for our simulations, we compiled all circuits in terms of the native gates specified in \Cref{subsec:device}.}
    \label{fig:aux_syndr_circ}
\end{figure*}

A simple QEC experiment, like quantum memory or stability, with a Calderbank-Shor-Steane (CSS) code is composed of three steps: initialising the data qubits of a code in either the $X$ or $Z$ basis, measuring a set of stabilisers some number of times, and finally measuring the data qubits in the $X$ or $Z$ basis. More complicated QEC experiments, like lattice surgery, are composed of more steps. The stabilisers may be measured using standard syndrome extraction circuits shown in \Cref{fig:aux_syndr_circ}. With standard circuits, the stabiliser measurements directly correspond to the outcomes of the auxiliary qubits. More complex circuits, such as circuits with flag qubits \cite{flag} or the circuit we discuss in \Cref{sec:swappy}, obtain further information.

In order to detect errors, we define detectors~\cite{McEwenBaconGidney} that are measurement combinations with deterministic values in the absence of errors. We explain the detector definitions for simple CSS-code QEC experiments, as outlined in the previous paragraph, that use standard syndrome extraction with unconditional resets, like \Cref{fig:aux_syndr_circ}(a). We may compose detectors for such experiments in the following way. For each stabiliser $g$, denote by $m_{g,j}$ the outcome from the $j$th QEC round, where $j\in\{1,2,\dots, n\}$, and let $(x,y)$ be the spatial coordinate of the auxiliary qubit of $g$. Between the first and last rounds of a QEC experiment, we define $m_{g,j}\oplus m_{g,j+1}$ as a detector at coordinate $(x,y,j)$ for all stabilisers $g$ and numbers of rounds $j\in\{1,2,\dots, n-1\}$. In the first round of QEC, if $g$ is of the same type as the data qubit initialisation, then we assign $m_{g,1}$ as a detector with coordinate $(x,y,0)$.  In the last round of QEC, for stabilisers of the same type as the data qubit measurement, we assign a detector at coordinate $(x,y,n)$ defined as $m_{g,n}\oplus\bigoplus_{q\in\mathrm{supp}(g)}m_q$, where $m_q$ is the measurement outcome of data qubit $q$. The first two coordinates of detectors are called spatial coordinates, while the last coordinate is called a time coordinate.

\begin{figure*}
    \includegraphics[width=0.9\textwidth]{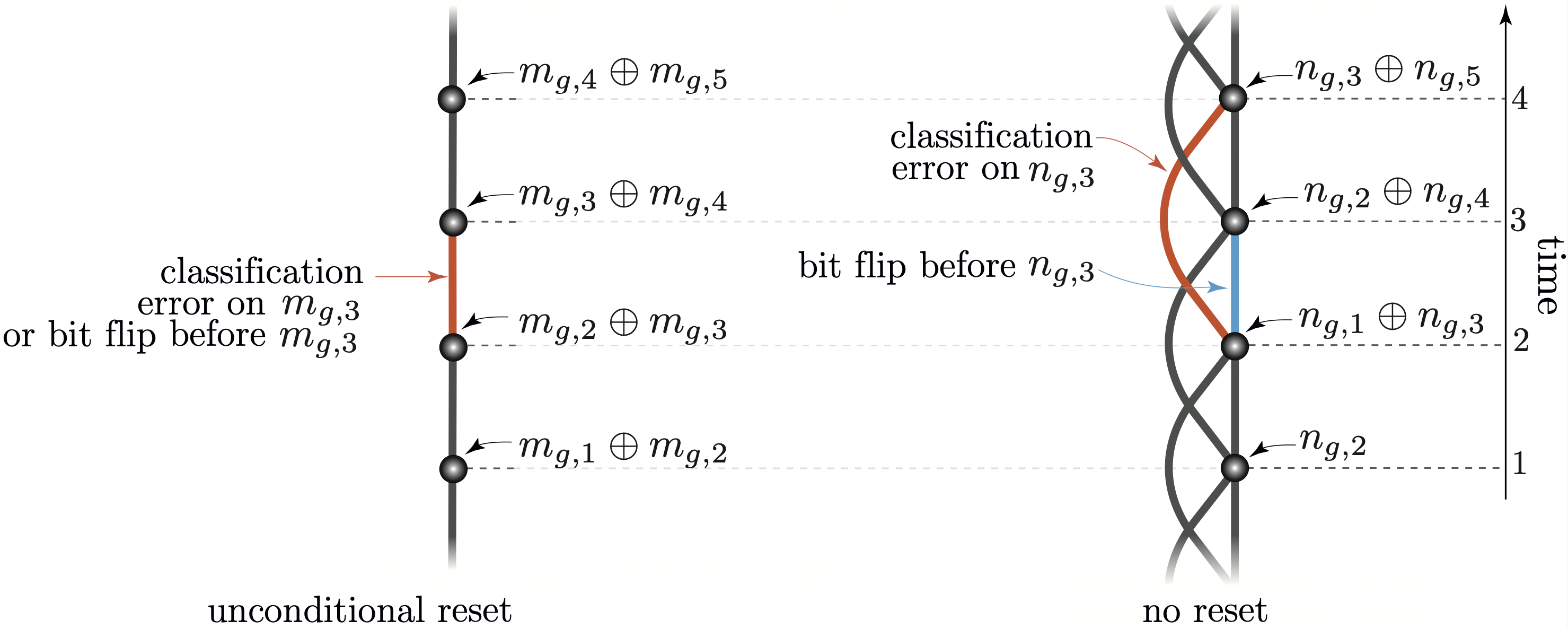}
    \caption{Decoding graphs for measurements from a single auxiliary qubit in the unconditional-reset (left) and no-reset (right) cases.}
    \label{fig:measerrors}
\end{figure*}

Now, we analyse which detectors are triggered (change value from the noise-free case) by different errors. A Pauli error on a data qubit occurring during the initialisation of the auxiliary qubits triggers only detectors with the same time coordinate. We refer to errors with this property as ``space-like''. On the other hand, a measurement error (due to either quantum bit flip or classical misclassification) on the auxiliary qubit of $g$ in the $j$th QEC round triggers at most two detectors, namely at coordinates $(x,y,j-1)$ and/or $(x,y,j)$, as shown on the left of \Cref{fig:measerrors}. Such an error is described as ``time-like''. Some Pauli errors that occur mid-syndrome-extraction-circuit trigger a set of detectors where neither the spatial nor the time coordinates coincide for all the detectors -- these are known as hook errors \cite{Dennis-Kitaev-Landahl}.

In order to evaluate the errors that have occurred during the experiment, the decoder takes in the triggered detectors along with a model of which errors trigger which detectors. For the planar code, in the case of a circuit-level Pauli noise model such as that in \Cref{subsec:device}, to a good approximation, this model can be captured in a graph and therefore graph-based decoders such as minimum-weight perfect matching (MWPM)~\cite{Dennis-Kitaev-Landahl,pymatching} can be used. We declare success in a run of the experiment if the decoder correctly predicts whether some logical observable has changed due to errors. In the case of $W$-quantum memory, for $W\in\{X,Z\}$, this logical observable is given by $\bigoplus_{q\in\mathrm{supp}(\widehat{W})} m_q$, where $\widehat{W}$ is the logical $W$ operator of the code.

We now consider the case where we apply a bit-flip gate to each auxiliary qubit after measurement conditioned on the outcome (\Cref{fig:aux_syndr_circ}(b-c)) instead of applying unconditional resets. We may apply these $X$ gates in hardware, to give the conditional-reset scheme (\Cref{fig:aux_syndr_circ}(b)), or track their effect in software, to give the no-reset scheme (\Cref{fig:aux_syndr_circ}(c)). As we will see, in both schemes, one measurement classification error leads to a pair of correlated errors, unlike with unconditional reset -- a previously overlooked phenomenon. In the conditional-reset case, we define the detectors as above; in the no-reset case, we determine the detectors as follows.

For the no-reset scheme, we calculate the effect of removing all conditional $X$ gates by pushing them through the subsequent unitary Clifford gates and find that a conditional $X$ can be accounted for by flipping the subsequent measurement~\cite{varbanov2020leakage}. Therefore, we can determine the stabiliser observable outcomes, $m_{g, j}$, from the no-reset measurement outcomes labeled $n_{g, j}$ as
\begin{align}
    m_{g,j} = n_{g,j-1}\oplus n_{g,j} \quad \text{for} \; j=1,2,\dots,n, \label{eq:m_expressed_with_ns}
\end{align}
where we define $n_{g,0}=0$; see \Cref{fig:aux_syndr_circ}(c). Between the first and last rounds of QEC, the detectors at coordinate $(x,y,j)$ are then
\begin{align}
   m_{g,j} \oplus m_{g,j+1} & = n_{g,j-1} \oplus n_{g,j+1} \quad \text{for} \; j=1,2,\dots,n-1, \label{eq:detectors_with_ns} 
\end{align}
where we have used $n_{g,j} \oplus n_{g,j}=0$ to simplify the expression.  Therefore, instead of comparing outcomes from consecutive rounds, we compare outcomes that are two rounds apart.  

Next, we consider the special case of the first and last QEC rounds. In the first QEC round, for each stabiliser $g$ of the same type as the data qubit initialisation, we assign $n_{g,1}$ as a detector with coordinate $(x,y,0)$. In the last round of QEC, for stabilisers of the same type as the data qubit measurement, we assign a detector at coordinate $(x,y,n)$ for $n_{g,n-1}\oplus n_{g,n}\oplus\bigoplus_{q\in\mathrm{supp}(g)}n_q$. 

This structural difference of detectors does not change the triggered detectors corresponding to any mid-circuit quantum Pauli error. However, misclassification of a measurement outcome results in a different combination of triggered detectors to the unconditional-reset case. To see this, let us assume we have a classification error on measurement result $n_{g,j}$, i.e., we read out $n_{g,j}\oplus 1$, even though the qubit collapsed into the $n_{g,j}$-eigenstate. This triggers two detectors at coordinates $(x,y,j-1)$ and $(x,y,j+1)$ (cf. \Cref{eq:detectors_with_ns}), so in the decoding graph it corresponds to a time-like edge with length $2$. In contrast, even in the no-reset case, a quantum measurement error corresponds to a time-like edge with length 1. A bit-flip on the auxiliary qubit associated with stabiliser $g$ just before measurement in the $k$th QEC round flips all measurements $n_{g, j}$ with $j \geq k$. This therefore triggers the detectors at coordinates $(x, y, k-1)$ and $(x, y, k)$. Both types of error are shown on the right of \Cref{fig:measerrors}.

In \Cref{fig:3D_graph_diagrams}, we present a topological perspective on how vertical strings of classification errors can lead to an undetectable logical failure in lattice surgery and stability experiments.  In a lattice surgery operation, we measure a joint logical Pauli between 2 or more logical qubits, and the result is determined from a (corrected) product of stabiliser measurement outcomes.  An uncorrected vertical string of classification errors will flip the value of one of these stabiliser measurements, and therefore flip the outcome of this logical Pauli measurement.  The need to suppress such failure modes is why $d$ QEC rounds is the standard recommendation. When lattice surgery is used to perform a non-Clifford gate (e.g. a $T$ gate), we apply a logical Clifford gate conditional on the outcome of the lattice surgery measurement result.  In such a situation, a logical measurement error during lattice surgery would be converted into a logical qubit error.  

Such lattice surgery circuits are large and complex. Fortunately, we can instead use the smaller and simpler stability experiment, in simulation or on real qubits,  to quantify the probability of logical failures due to a vertical string of errors. Stability experiments are therefore an excellent proxy for lattice surgery operations. Considering these vertical failure mechanisms, we see that the no-reset approach will tolerate only half as many classification errors as the unconditional-reset approach, since each error has twice the vertical length in the former scheme.  This insight applies equally to lattice surgery and stability experiments.   As we will see in \Cref{subsec:stability}, this difference significantly impacts the performance of the stability experiment, and hence lattice-surgery-based FTQC.  For other types of logical operation such as transversal gates, logical failure modes are less well understood -- there is no known proxy in the same spirit as stability -- and the required number of QEC rounds is an ongoing topic of debate  \cite{sahay2024error,wan2024iterative,zhou2024algorithmic} that we regard as unsettled.    For quantum memory, the no-reset approach does not reduce the weight of minimal-weight logical errors, and so in this context we should expect only a minor variation in logical error rates.

In the conditional-reset scheme, a classification error on $m_{g, j}$ further results in erroneously applying an $X$ gate after measurement. As a result, the next measurement outcome $m_{g, j+1}$ is also flipped. Thus, two detectors are triggered at $(x,y,j-1)$ and $(x,y,j+1)$, introducing a time-like edge with length $2$, as in the no-reset case.

\section{QEC experiments with standard syndrome extraction circuits}\label{sec:qec_experiments}

\subsection{Noise model}\label{subsec:device}
This work is motivated by the absence of unconditional resets in recent QEC experiments on superconducting devices~\cite{Delft-13, ETH-planar-code}. Therefore, the numerical results presented in this paper are from simulating a quantum computer with superconducting-like properties. These properties are the qubit connectivity, the native gates and the noise model.

We assume that the device has fixed and uniform qubit connectivity, with a planar nearest-neighbour architecture connecting each (bulk) qubit to four others. This so-called square-grid connectivity enables implementation of the planar code and is found in several modern superconducting quantum computers~\cite[e.g.,][]{Rigetti-Ankaa, acharya_suppressing_2023}. In \Cref{sec:swappy}, we will present a circuit that requires slightly different planar connectivity -- some qubits need only be connected to three others instead of four.

\begin{table}[]
\footnotesize
   \centering
    \begin{tabular}{|c|c|}
        \hline
        Operation & Duration in ns \\
        \hline
        $\sqrt{X}$, $S$ gate& 20 \\
        CZ gate & 40 \\
        measurement & 600 \\
        (unconditional) reset & 500 \\
        reference $T_{1}$ & 30,000 \\
        reference $T_{2}$ & 30,000\\
        \hline
    \end{tabular}\\
    \vspace{2mm}
    \begin{tabular}{|c|c|}
        \hline
        Error mechanism & Probability \\
        \hline
        1Q depolarisation after $\sqrt{X}$ or $S$ & $p$/10 \\
        2Q depolarisation after CZ & $p$ \\
        1Q bit-flip after reset & 2$p$ \\
        1Q bit-flip before measurement & 4$p$ \\
        classical measurement flip & $p$ \\
        \hline
    \end{tabular}
    \caption{Properties of our noise model inspired by current superconducting devices. (Top) Native qubit operations with their duration in nanoseconds, and reference $T_1$ and $T_2$ times corresponding to physical error rate $p=0.01$. As $p$ changes, the $T_1$ and $T_2$ times are scaled accordingly (see main text). In \Cref{subsec:stability}, we also consider faster resets; the $500$~ns value here corresponds to slow reset. (Bottom) The noise channels associated with each operation. Notably, the measurement error has quantum and classical components that have the same effect in the unconditional-reset scheme; however, their effect is different in the no-reset (or conditional-reset) scheme.}
    \label{tab:timesandnoise}
\end{table}

We construct our simulated circuits \cite{our_stim_circuits} in terms of the controlled-$Z$ two-qubit gate (CZ) and single-qubit $X$- and $Z$-basis rotations, a common native gate set for superconducting qubits~\cite[e.g.,][]{Rigetti-Ankaa}. In particular, we use the $\pi/2$ rotations $\sqrt{X}$ and $S$, given by
\begin{equation}
    \sqrt{X} = \frac{1}{\sqrt{2}}(I - i X), \quad S = \frac{1}{\sqrt{2}}(I - i Z).
\end{equation}
We also use $Z$-basis measurement (MZ) and reset (RZ). Each of these operations has an associated duration, given in \Cref{tab:timesandnoise}. In particular, measurement and reset take an order of magnitude longer than one- and two-qubit gates (see also \cite{Delft-13,Delft-planar-docde}). The CZ gate takes only twice as long as a one-qubit gate.

We parameterise our noise model with a single parameter, $p$. The $p$-dependent probabilities of different error mechanisms are given in \Cref{tab:timesandnoise}. These probabilities are based on an existing superconducting-inspired noise model~\cite{gidney2021fault}, with a few small differences. Firstly, we split the measurement noise into two -- a quantum part, which is bit-flip noise applied to the qubit before the measurement, and a classical part, which is the misclassification of the outcome value, e.g., reading out $0$ even though the qubit collapsed into the $|1\rangle$ state. This distinction is not commonly made, as they have the same effect under the unconditional-reset scheme. However, with no reset this distinction becomes important as the two types of measurement noises have different effects. We assign $80\%$ and $20\%$ of the noise respectively to these two mechanisms, inspired by existing hardware properties~\cite{Delft-13}. Secondly, when a qubit is idling, we apply noise depending on the length of time for which it is idle, $t$. The noise is obtained by Pauli twirling the amplitude damping and dephasing channel to give a Pauli channel with associated probabilities
\begin{align}
    p_{X}(t) &= p_{Y}(t) = \frac{1}{4} \left(1 - e^{-t/T_1} \right), \\
    p_Z(t) &= \frac{1}{2} \left(1 - e^{-t/T_2} \right) - \frac{1}{4} \left(1 - e^{-t/T_1} \right),
\end{align}
where $p_{W}$ is the probability of a Pauli-$W$ error occurring~\cite{sarvepalli2009asymmetric}. To parameterise the idle noise by $p$, we scale the $T_{1}$ and $T_{2}$ times so that
\begin{equation}
    T_{1} = \frac{p^{\mathrm{ref}}}{p}T_{1}^{\mathrm{ref}}, \quad T_{2} = \frac{p^{\mathrm{ref}}}{p}T_{2}^{\mathrm{ref}}
\end{equation}
where $T_{1}^{\mathrm{ref}}$ and $T_{2}^{\mathrm{ref}}$ are given in \Cref{tab:timesandnoise} and $p^{\mathrm{ref}} = 0.01$.

We note that our chosen noise model is not meant to capture the exact behaviour of any specific device. The analysis could be repeated with different device assumptions which may change the quantitative details of the results. However, it captures the key features of superconducting hardware and will be qualitatively representative of other qubit types.

\subsection{The impact of no-reset on memory}\label{subsec:qmem}

\begin{figure}[t!]
        \centering
        \includegraphics[width=0.49\textwidth]{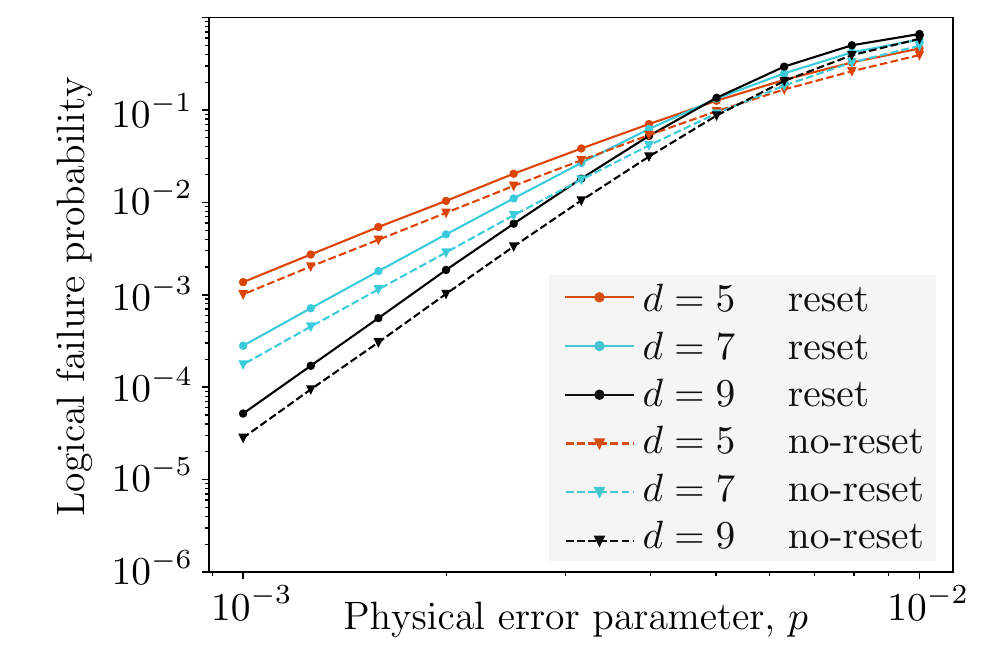}
    \caption{Comparison of quantum memory performances in the unconditional-reset (dots and solid lines) and no-reset (triangles and dashed lines) schemes using standard syndrome extraction circuits. For each distance $d = 5, 7, 9$ (corresponding to red, blue, black colours), we performed $d$ QEC rounds. The numbers of samples taken to calculate the data points are sufficiently large for the error bars not to be visible. As can be seen, there is only a small difference between the two schemes, confirming the statement made in \cite[Sec. S2]{miao_overcoming_2022}.   Note that the conditional reset scheme is not shown as the simulation is identical to the no-reset simulations except with higher noise levels due to the increased physical qubit idling. As a consequence, the conditional-reset scheme will always perform worse and is therefore not shown here.}
    \label{fig:mem_no_gadget_threshold}
\end{figure}

Using the noise model described in \Cref{subsec:device}, we performed quantum memory simulations for distance-$d$ rotated planar code for $d$ number of QEC rounds. \Cref{fig:mem_no_gadget_threshold} compares the unconditional-reset and no-reset schemes. We constructed circuits using the python library \texttt{stim} \cite{stim} and took samples which were then decoded using the MWPM python library \texttt{pymatching} \cite{pymatching}. The circuits are available here: \cite{our_stim_circuits}. The number of samples for each data point was the minimum required to observe $10^4$ logical failures or reach $10^8$ samples, whichever happened first. This ensures error bars that are too small to be visible in our plots. We calculated the sample logical failure probabilities $p_X$ and $p_Z$, for $X$ and $Z$-memory, respectively, and combined them into one quantity as $p_L = p_X+p_Z-p_Xp_Z$. Our results (\Cref{fig:mem_no_gadget_threshold}) confirm the statement of \cite{miao_overcoming_2022}, mentioned earlier, that there is only a small difference between the two reset schemes in the context of quantum memory. More precisely, the no-reset scheme's performance is slightly better, as the threshold is somewhat higher and the logical failure probabilities are slightly lower. One reason for this difference is that the idle noise on the data qubits is lower in the no-reset case.

Note, however, that, as we will see in \Cref{subsec:stability}, the time-like effective distance (i.e., the number of mid-circuit errors that cause an undetectable time-like logical failure) is halved in the no-reset scheme for the stability experiment. Therefore, in the no-reset scheme, it may be more reasonable to perform $2d-1$ number of QEC rounds instead of $d$. Increasing the number of rounds for quantum memory degrades its performance, eroding the slight advantage of the no-reset scheme.

\begin{figure*}[t!]
    \centering
    \includegraphics[width=0.80\textwidth]{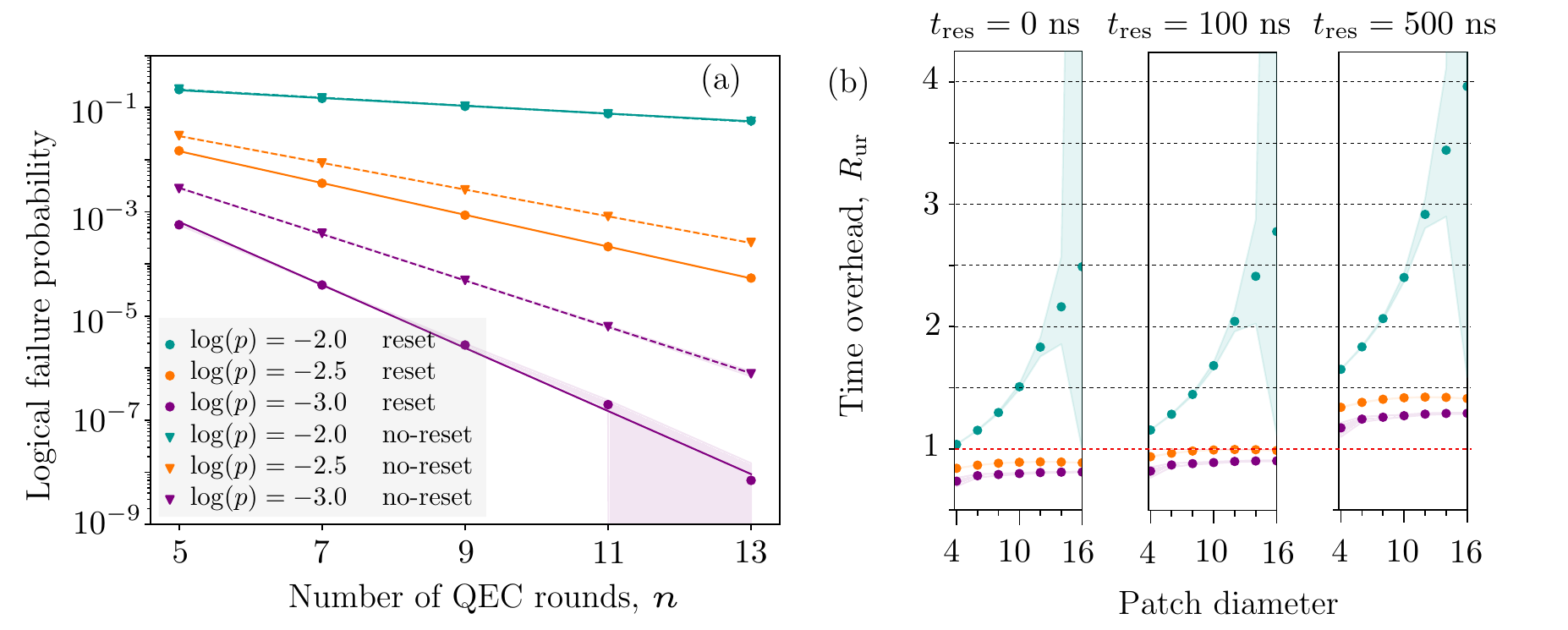}
    \caption{Comparison of the stability experiment performances in the two reset schemes using standard syndrome extraction circuits. Colours indicate different physical error probabilities: $p = 10^{-2}$ (blue), $p = 10^{-2.5}$ (orange), $p = 10^{-3}$ (purple).
    (a) Logical failure probabilities of the stability experiment for the $4\times 4$ patch shown in \Cref{fig:planar_code}(b) plotted against the number of QEC rounds. Error bars corresponding to $3\times$ the standard error of the mean are also shown as shaded areas. Note that for the $p = 10^{-2}$ case the plots overlap. The unconditional reset is plotted with dots, the no-reset with triangles. Best $\log$-line fits are also shown with solid and dashed lines, respectively. 
    (b) The time overhead $R_{\mathrm{ur}}$ that the unconditional-reset scheme requires when compared to the no-reset scheme given $t_{\mathrm{res}}=0$ (instantaneous), $100$ (fast) and $500$~ns (slow) resets, respectively. In particular, the first column of dots in the first plot corresponds to (a), since instantaneous reset implies equal execution time in the two reset schemes. The shaded areas correspond to $90\%$ confidence intervals. The pink dashed line shows the break-even point, i.e. when $R_{\mathrm{ur}}=1$, below which the unconditional reset has better FTQC performance.  Note that the conditional reset scheme is not shown as the simulation is identical to the no-reset simulation except with higher noise levels due to the increased physical qubit idling. As a consequence, the conditional reset scheme will always perform worse and is therefore not shown here.}
    \label{fig:stability_no_gadget_plots}
\end{figure*}

\subsection{The impact of no-reset on FTQC}\label{subsec:stability}

During FTQC, there are purely time-like undetectable logical failures that would be affected by the choice of reset scheme. For instance, if we measure a joint logical Pauli product via lattice surgery between some logical qubits that are each encoded into a planar code, then this outcome is given as a joint parity of a set of stabilisers \cite{Horsman_2012, ChamCamtwistfree, ChamCamtwistbased}; see also \Cref{fig:3D_graph_diagrams}(c). To make this fault-tolerant, we measure these stabilisers for $n$ QEC rounds, with $n$ sufficiently large. With the unconditional-reset scheme, the minimum number of errors that causes an undetectable logical failure is $n$, e.g., when one stabiliser is misclassified in all QEC rounds. However, with the no-reset scheme, this number becomes $\lceil n/2 \rceil$, as now it is enough to misclassify this measurement in the first, third, fifth, etc. QEC rounds.

The stability experiment \cite{Gidney-stability} captures this type of situation in a simplified proxy and thus can be used to assess the necessary number of QEC rounds during lattice surgery and other operations where time-like logical failures are relevant. We use a planar code patch that does not encode any logical qubits, but, instead, has one type of stabiliser that is over-determined. \Cref{fig:planar_code}(b) shows an example $4\times 4$ stability patch where the $X$-type stabilisers multiply into the identity operator and so are over-determined. This stability experiment is performed as follows: we initialise all data qubits in the $|0\rangle$ state, then measure the stabilisers for $n$ rounds, and finally we measure all data qubits in the $Z$ basis. We assign detectors as in \Cref{sec:detectors_no_reset}, and define the logical observable as the product of all $X$-type stabilisers in the first QEC round. Mirroring lattice surgery, a minimum of $n$ measurement errors amount to an undetectable logical failure in the unconditional-reset scheme, while $\lceil n/2 \rceil$ classification errors are sufficient in the no-reset scheme, cf. \Cref{eq:m_expressed_with_ns,eq:detectors_with_ns}.

Based on this, we would expect the stability performance to be improved by using unconditional resets -- assuming fast enough and high-enough fidelity reset gates. We performed simulations to assess this. We considered planar code stability patches of sizes $w \times w$ with $w \in\{4,6,8,10,12,14,16\}$ and physical error rates $p\in\{10^{-2},10^{-2.5},10^{-3}\}$. For each, we prepared \texttt{stim} circuits \cite{our_stim_circuits}, sampled, and decoded until we reached either $10^9$ total shots or $10^6$ logical failures, whichever happened first. The obtained logical failure probabilities plotted against the number of QEC rounds, $n=5, 7, 9, 11, 13$, are shown in \Cref{fig:stability_no_gadget_plots}(a) for the $w=4$ case. They clearly show that implementing unconditional reset improves the performance for $p = 10^{-2.5}, 10^{-3}$. Furthermore, this improvement increases as $p$ decreases. 

We now quantify this improvement. In the stability experiment, we expect the logical failure probability $p_L$ to satisfy
\begin{equation}\label{eq:lg_pl_against_n}
    \log(p_L) = \log(a) - \gamma n,
\end{equation} where $a$ and $\gamma$ both depend on the width of the patch $w$ \cite[Section 3]{Gidney-stability}. The $\gamma$ parameter quantifies the strength of the exponential error suppression with the number of rounds in the stability experiment.
Note that one QEC round takes $t_{\mathrm{nr}} = 840$~ns with the no-reset scheme and, if the reset duration is $t_{\mathrm{res}}$~ns, one QEC round takes $t_{\mathrm{ur}}=(840+t_{\mathrm{res}})$~ns with the unconditional-reset scheme. These two times are equal only with instantaneous reset, i.e., $t_{\mathrm{res}}=0$~ns. Therefore, as a fairer comparison, \Cref{eq:lg_pl_against_n} may be expressed in terms of the time $t$ the experiment requires, instead of the number of QEC rounds, so that
\begin{align}\label{eq:lg_pl_against_t_nr_and_wr}
    \log(p_{\mathrm{nr}, L}) & = \log(a_{\mathrm{nr}}) - \gamma_{\mathrm{nr}} t \\
    \log(p_{\mathrm{ur}, L}) & = \log(a_{\mathrm{ur}}) - \gamma_{\mathrm{ur}} t,
\end{align} 
where subscripts $\mathrm{nr}$ and $\mathrm{ur}$ indicate the no- and unconditional-reset schemes, respectively. We define the time overhead of implementing unconditional resets as $R_{\mathrm{ur}}=\gamma_{\mathrm{nr}}/\gamma_{\mathrm{ur}}$ \cite[Sec. 5.2]{geher_tangling_2023}. This can be interpreted as follows: if we perform a stability experiment for $t_{\mathrm{nr}}$~ns without unconditional resets then, in order to match this performance using unconditional resets, we need to spend approximately $t_{\mathrm{ur}} = R_{\mathrm{ur}}t_{\mathrm{nr}}$~ns. If $R_{\mathrm{ur}}<1$, then implementing unconditional resets decreases the time cost of FTQC, and hence is worth implementing; otherwise not. The time overhead is plotted in \Cref{fig:stability_no_gadget_plots}(b), assuming instantaneous reset $t_{\mathrm{res}} = 0$~ns, fast reset $t_{\mathrm{res}} = 100$~ns, and slow reset $t_{\mathrm{res}} = 500$~ns. The plotted error bars correspond to $90\%$ confidence intervals. More precisely, for each data point, we sampled $1000$ times from a normal distribution with mean given by the sampled logical error probability, and standard deviation given by the standard error of the mean. Then, we took the best $\log$-line fits for each of the $1000$ cases, calculated the corresponding time overheads, and removed the $50$ smallest and the $50$ largest values hence obtained. The minimum and maximum of the remaining values are the bottom and top error bars, respectively.  We call $p_{\mathrm{br}}$ the break-even point, where $R_{\mathrm{ur}}=1$. Clearly, as $t_{\mathrm{res}}$ increases, $p_{\mathrm{br}}$ decreases. Furthermore, it can be seen in \Cref{fig:stability_no_gadget_plots}(b) that, as $p$ decreases, the time overhead decreases too, indicating that when $p$ is small enough, implementing unconditional reset improves FTQC performance. In all cases of instantaneous and fast reset, the break-even point satisfies $10^{-2.5} < p_{\mathrm{br}} < 10^{-2}$. With slow reset, the break-even point is not visible in our plot, and we expect it to be at very low $p$. This indicates that, unless the reset gate error rate and duration are below certain values, implementing unconditional reset may not be beneficial for FTQC. However, if the break-even point is reached, either with decreased reset duration or higher-fidelity reset gates, the unconditional-reset scheme substantially improves FTQC performance.

\section{Recovering time-like distance by spreading classification errors}\label{sec:spread}
We have seen that the absence of unconditional resets halves the time-like effective distance of FTQC protocols when using standard syndrome extraction circuits. In this section, we present an alternative syndrome extraction circuit that recovers the full time-like effective distance without the need for unconditional resets, and which also keeps the space-like effective distance as $d$. This circuit uses one additional two-qubit gate per QEC round and an additional $O(d)$ qubits.

\begin{figure}[t!]
\centering
 \includegraphics[width=0.3\textwidth]{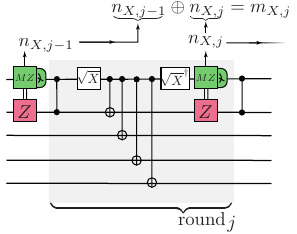}
    \caption{Syndrome extraction circuit for an $XXXX$ stabiliser that spreads the classification error to a data qubit. Immediately after measurement, we apply a classically-controlled phase-flip and a CZ quantum gate. There is no overall effect, unless there was a classification error on $n_{X,j-1}$, in which case a $Z$ error is transferred to a data qubit.}
    \label{fig:spread_gadget_circuits}
\end{figure}

\begin{figure}
\centering
    \includegraphics[width=0.4\textwidth]{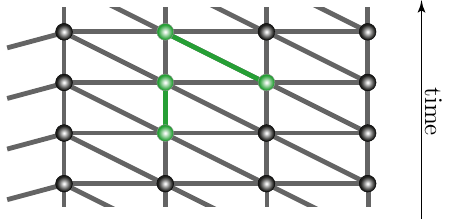}
    \caption{A cross-section of the decoding graph for the error-spreading circuit. In the bulk, it is the same as for standard unconditional-reset circuits. A classification error triggers four detectors (green nodes). This error can be decomposed into a quantum measurement error and a hook error (green edges).}
    \label{fig:graphs}
\end{figure}

\begin{figure}
\centering
    \includegraphics[width=0.3\textwidth]{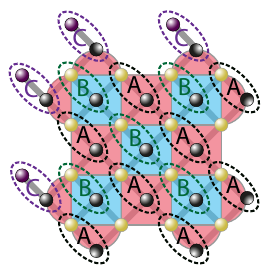}
    \caption{A $4\times 4$ stability patch with additional qubits (purple) that are needed for the error-spreading circuit. Pairs of qubits that take part in each error-spreading event are circled. Type A (black) and B (green) are auxiliary and data qubit pairs for $X$- and $Z$-type stabilisers, respectively, that do not require additional qubits. Type C (purple) are auxiliary and additional qubit pairs for the top and left boundary $X$-type stabilisers. For these, we measure a weight-three stabiliser and, at the end of the QEC round, we measure out the additional qubits simultaneously with the auxiliary qubits.}
    \label{fig:gadgetpatch}
\end{figure}

The idea is to spread the auxiliary qubit measurement classification errors to the data qubits, thereby triggering additional detectors and thus requiring more errors for an undetectable time-like logical failure. We achieve this by applying two conditional Pauli gates immediately after measuring the auxiliary qubit -- one conditioned on the measurement outcome and the other conditioned on the auxiliary qubit state itself; see \Cref{fig:spread_gadget_circuits} for an $X$-stabiliser circuit. Note that the controlled gates are of $Z$-($X$-)type for $X$-($Z$-)type stabilisers, and that the classically-controlled gates do not have to be applied on the device -- instead, their effect can be tracked in software. In the absence of a measurement classification error, the two conditional gates cancel so there is no overall effect on the data qubits. However, if there is a classification error, exactly one of the gates will have an effect, resulting in a Pauli gate being applied to the data qubit which then behaves like a Pauli error on the qubit. Therefore, misclassification of measurement $n_{g, j}$ will now trigger four detectors -- the two which directly result from the measurement classification error, at coordinates $(x, y, j-1)$ and $(x, y, j+1)$, and the two which arise from the resulting data qubit error, at coordinates $(x, y, j)$ and $(x', y', j)$. This pattern is shown in \Cref{fig:graphs}, together with its decomposition into two graph-like errors. Reference \cite{our_stim_circuits} contains an example decoding graph for a $4\times 4$ stability experiment using the error-spreading circuit that we generated with \texttt{stim}. In \Cref{fig:gadgetpatch}, we show a $4\times 4$ planar code stability patch and the qubit pairs which take part in error-spreading events. In the bulk of the code, we require no additional qubits. However, on the boundary, we require $O(w)$ additional qubits for a $w\times w$ patch to ensure each auxiliary qubit has a qubit to which to spread its classification error; cf. \Cref{fig:planar_code}(b).

It is easy to see that an undetectable logical error will no longer occur if there are classification errors in only every other round because of the additional detectors now triggered by a classification error. 
Furthermore, by searching exhaustively with \texttt{stim}, we find that, at least for small examples, $n$ errors are required for an undetectable logical failure in an $n$-round stability experiment. For quantum memory, it is straightforward to see that we need at least $d$ errors for an undetectable logical failure, as each classification error only triggers detectors whose spatial coordinates are adjacent, as in the standard circuit case.

\begin{figure}[t!]
    \centering
        \includegraphics[width=0.48\textwidth]{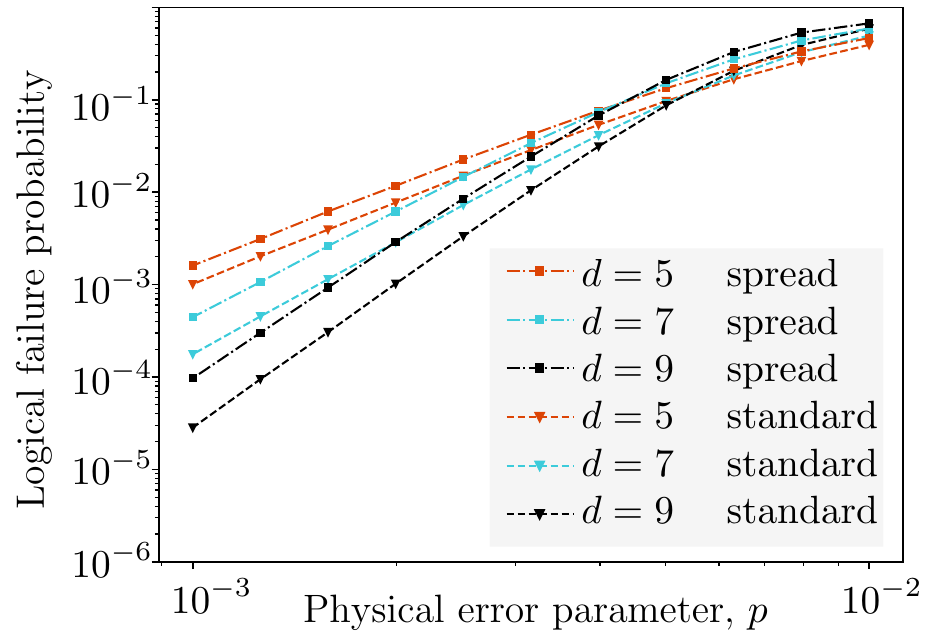}
    \caption{Quantum memory performance of rotated planar codes in the no-reset scheme using standard (triangles and dashed lines) and error-spreading (squares and dash-dotted lines) circuits. For each distance $d = 5,7,9$ (corresponding to red, blue, black colours), we performed $d$ QEC rounds.}
    \label{fig:mem_threshold_spread_gadget}
\end{figure}

\begin{figure}[h!]
    \centering
        \includegraphics[width=0.48\textwidth]{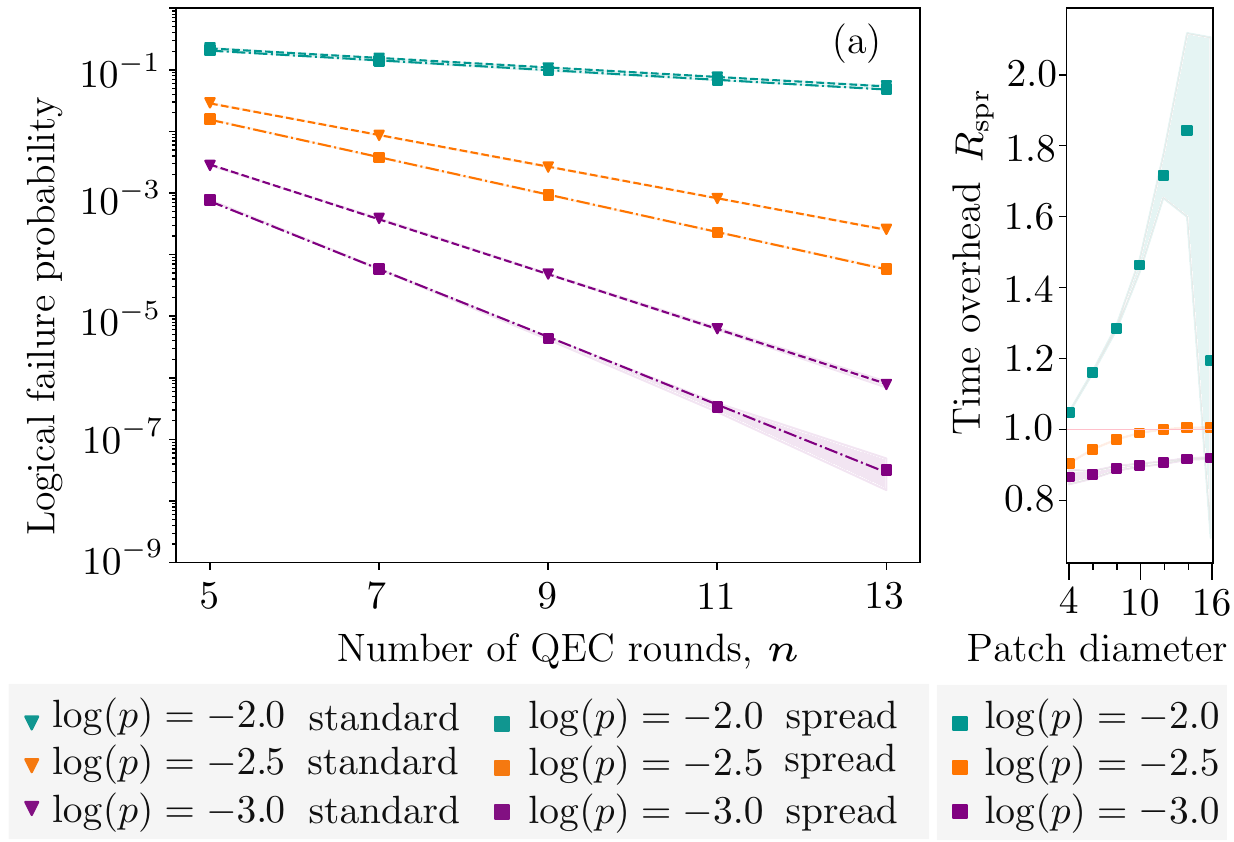}
    \caption{Comparison of stability experiment performances in the no-reset scheme using standard (triangles and dashed lines) and error-spreading (squares and dash-dotted lines) circuits. Colours indicate different physical error probabilities: $p = 10^{-2}$ (blue), $p = 10^{-2.5}$ (orange), $p = 10^{-3}$ (purple). (a) Comparison of logical failure probabilities for the $4\times 4$ stability patch. Error bars as in \Cref{fig:stability_no_gadget_plots}(a) are also shown as shaded areas. Note that for the $p = 10^{-2}$ case, the two lines overlap.  (b) The time overhead $R_{\mathrm{spr}}$ that the error-spreading circuit requires when compared to the no-reset standard circuit. The shaded areas correspond to $90\%$ confidence intervals. The conditional reset scheme is not shown as the simulation is identical to the no-reset except with higher noise levels due to the increased physical qubit idling. As a consequence, the conditional reset scheme will always perform worse and is therefore not shown here.}
    \label{fig:stability_spread_gadget_plots}
\end{figure}

We simulated the error-spreading circuits in order to compare the performance to standard syndrome extraction circuits with no reset. We applied the same principles for the number of shots as in \Cref{sec:qec_experiments}. We present the quantum memory results in \Cref{fig:mem_threshold_spread_gadget} and see that the performance is somewhat (although not significantly) diminished from the standard no-reset case. We attribute this to the error-spreading feature and the slightly higher depth of the error-spreading circuits. However, as we see from the stability time overhead plots \Cref{fig:stability_spread_gadget_plots}(b), we need fewer QEC rounds for the error-spreading circuit to avoid undetectable time-like logical failures and so we need fewer QEC rounds for quantum memory as well, making the gap in \Cref{fig:mem_threshold_spread_gadget} smaller. For instance, in case of the $4\times 4$ patch for $p=0.001$, we have $R_{\mathrm{spr}}\approx 0.85$, meaning we have to spend only $85\%$ of the time on stabiliser measurements as with the standard no-reset circuits. Taking into account the differing execution times, we see that we need only $\approx 80\%$ of the number of QEC rounds with the error-spreading circuit. Based on this, we can conclude that the error-spreading circuit improves the QEC performance over the standard no-reset circuits.

From comparing \Cref{fig:stability_spread_gadget_plots}(b) and \Cref{fig:stability_no_gadget_plots}(b), we can further conclude that the error-spreading circuit is a good alternative to unconditional reset if the reset duration exceeds $100$~ns for the error regimes we considered. However, if the reset duration is decreased further, then unconditional reset outperforms the error-spreading circuit.

\section{Recovering time-like distance by squeezing two QEC rounds into one}\label{sec:swappy}

\begin{figure}[t!]
    \centering
        \centering
        \includegraphics[width=0.35\textwidth]{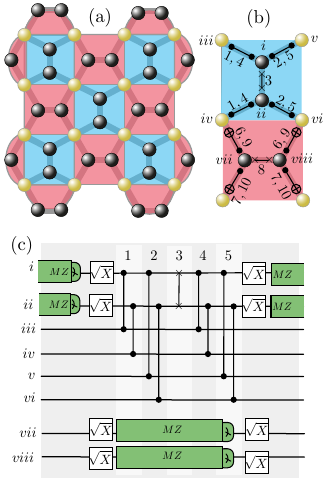}
    \caption{(a) A $4\times 4$ stability patch (cf. \Cref{fig:planar_code}(b)) for the round-squeezing circuit implementation laid out on the Cairo pentagonal connectivity that avoids ``bad hooks''. Each stabiliser has two auxiliary qubits as shown. (b) Graphical representation of the round-squeezing circuit for two adjacent stabilisers. In steps $1$--$5$, we apply CZ and SWAP two-qubit gates for the $Z$-type stabiliser as shown, during which we measure the auxiliary qubits of the $X$-stabiliser. Then, in steps $6$--$10$, we apply CX and SWAP two-qubit gates for the $X$-type stabiliser and, at the same time, measure the auxiliary qubits of the $Z$-type stabiliser. In the bulk of the experiment, we repeat this. In this way, the execution time of the round-squeezing circuit for the same number of QEC rounds is only slightly increased compared to the standard circuit's execution time with no reset, and is shorter than the execution time of standard circuits with unconditional reset, provided the reset duration is $>160$~ns. (c) Circuit diagram of steps $1$--$5$.}\label{fig:swappy_gadget}
\end{figure}

In this section, we show that QEC performance can be improved without using unconditional resets by effectively squeezing two QEC rounds into one. This method, however, requires $\approx 50\%$ additional qubits. We appoint two auxiliary qubits for each stabiliser, as shown in \Cref{fig:swappy_gadget}(a). Note that the auxiliary qubits of the $X$-/$Z$-type stabilisers are arranged horizontally/vertically so that we avoid so-called ``bad hook'' errors. Therefore, the hardware needs the Cairo pentagonal connectivity~\cite{Gidney-pentagonal}. The main idea is that while we measure the auxiliary qubits of one stabiliser type, we can execute unitary gates on the qubits of the other type and also have plenty of time to obtain two independent measurement outcomes for each stabiliser. \Cref{fig:swappy_gadget}(b)--(c) illustrates this. For instance, the unitary parts of this circuit for a $Z$-type stabiliser are as follows. We use Roman numerals for qubit labels and numerical labels for protocol steps.  First, apply two layers of CZ gates: step (1) $CZ_{i, iii}$ and $CZ_{ii, iv}$; then step (2)  $CZ_{i, v}$ and $CZ_{ii, vi}$. Then, swap the auxiliary qubits, so step (3) $SWAP_{i,ii}$.  Finally, repeat the two layers of CZ gates: step (5) $CZ_{i, iii}$ and $CZ_{ii, iv}$; then step (6)  $CZ_{i, v}$ and $CZ_{ii, vi}$. In this way, we entangle the two auxiliary qubits with the data qubits independently and so measuring each provides independent stabiliser measurements. Since executing these unitary gates takes only $400$~ns in total (when compiled to the gates of \Cref{tab:timesandnoise}), one QEC round with this circuit takes $1000$~ns. This is only $160$~ns longer than for the standard no-reset syndrome extraction; however, we obtain each stabiliser outcome twice.

\begin{figure}[t!]
    \centering
    \includegraphics[width=0.48\textwidth]{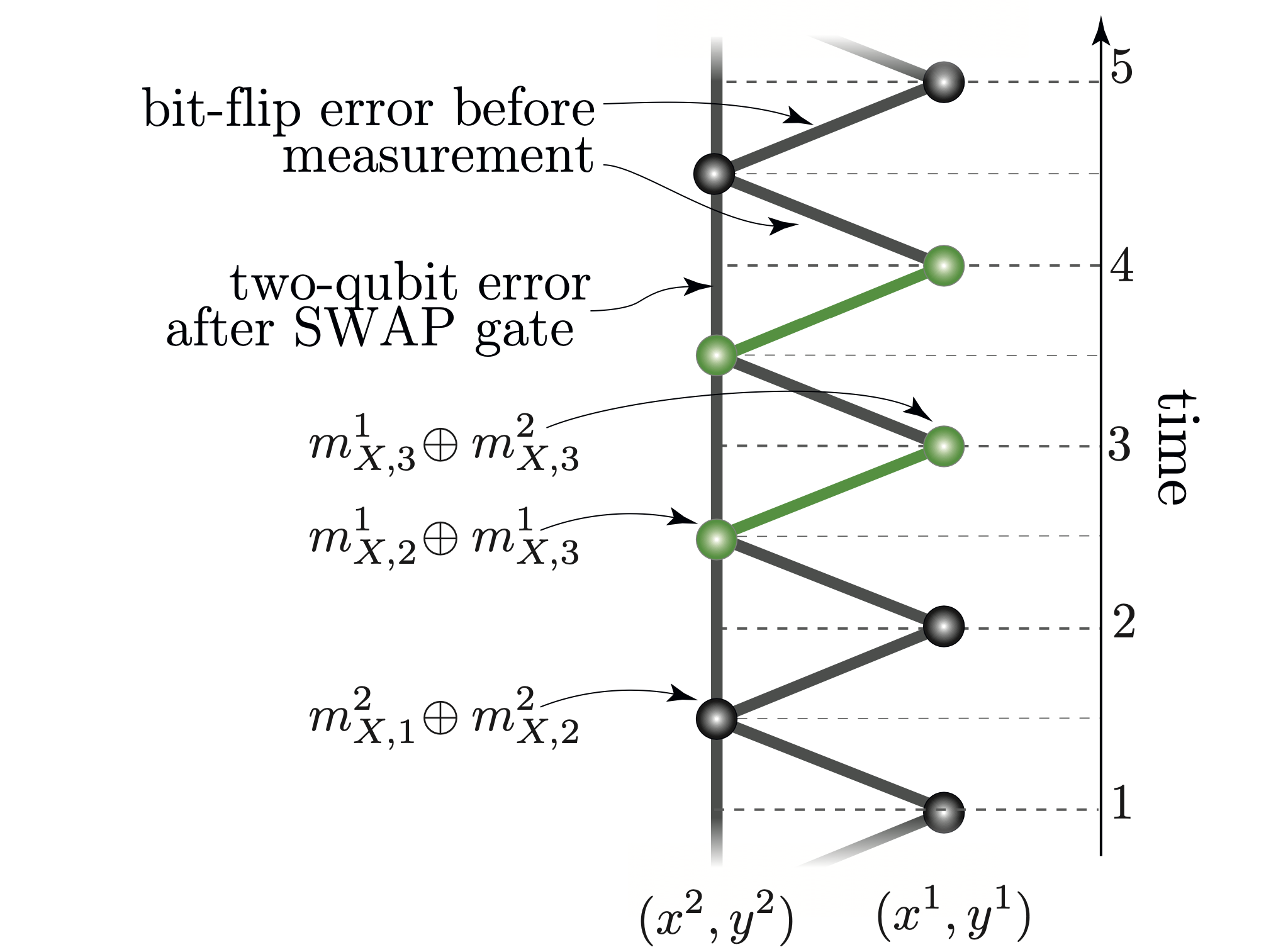}
    \caption{Part of the decoding graph for a stability experiment with the round-squeezing circuit that only contains nodes associated with one $X$-type stabiliser. A quantum bit-flip error before measurement corresponds to a diagonal edge, while an $XX$-error on the auxiliary qubits after the SWAP gate corresponds to a vertical edge. The measurement classification error triggers four detectors (green nodes) and can be decomposed into two diagonal edges (green edges).}\label{fig:swappy_decoding_graph}
\end{figure}

We now show how this new circuit recovers the effective time-like distance. First, we define the detectors. For each stabiliser $g$, let $(x^1,y^1)$ and $(x^2,y^2)$ be the spatial coordinates of its two auxiliary qubits. For the sake of simplicity, we will consider outcomes in the conditional-reset scheme, which is equivalent to the no-reset scheme. Denote by $m_{g,j}^k$ the outcome in the $j$th QEC round obtained on the auxiliary qubit at $(x^k,y^k)$ ($k=1,2$). We set $m_{g,j}^1 \oplus m_{g,j}^2$ as a detector for all $j$ and assign the coordinates $(x^1,y^1,j)$ to it. Furthermore, in the bulk of the experiment, we define $m_{g,j}^1 \oplus m_{g,j+1}^1$ as a detector if $j$ is even, and $m_{g,j}^2 \oplus m_{g,j+1}^2$ when $j$ is odd, and assign the coordinates $(x^2,y^2,j+\tfrac{1}{2})$ to it. The full decoding graph of a stability experiment on a $4\times 4$ patch with $5$ rounds is available in \cite{our_stim_circuits}. \Cref{fig:swappy_decoding_graph} depicts the part of the decoding graph where only nodes (detectors) corresponding to one $X$-type stabiliser are shown. This restricted graph has edges of two types: diagonal and vertical edges. If $m_{g,j}^k$ ($k\in\{1,2\}$) is flipped due to a quantum bit-flip measurement error, then at most two detectors are triggered, namely at coordinates $(x^1,y^1,j)$ and $(x^2,y^2,j\pm\tfrac{1}{2})$, corresponding to a diagonal edge. If $m_{g,j}^k$ ($k\in\{1,2\}$) is flipped due to misclassificiation, the most dangerous error for FTQC, then that will also flip $m_{g,j+1}^{3-k}$, hence triggers at most four detectors. However this can be decomposed into two parallel diagonal edges. To account for the vertical edges, consider an $XX$-error occurring after the SWAP gate in \Cref{fig:swappy_gadget}(b). This error flips both outcomes of the $j$th QEC round, hence triggers at most two detectors at coordinates $(x^2,y^2,j-\tfrac{1}{2})$ and $(x^2,y^2,j+\tfrac{1}{2})$ -- a vertical edge. We verified using \texttt{stim}'s functionality that the effective distance of stability experiments is recovered to be $n$. Furthermore, the effective distance for quantum memory remains the same as with standard syndrome extraction circuits: $d$.

\begin{figure}[h!]
    \centering
    \includegraphics[width=0.45\textwidth]{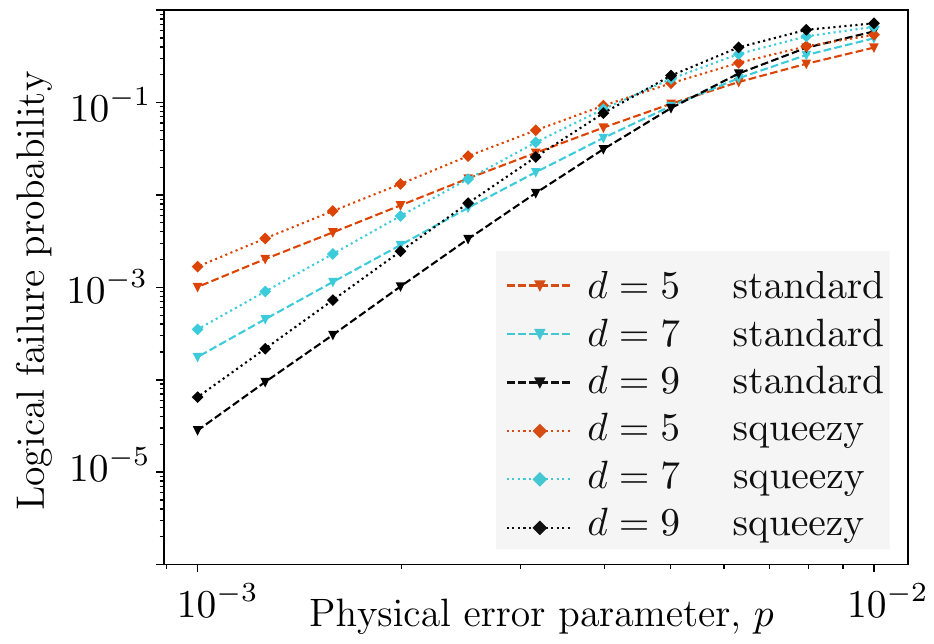}
    \caption{Comparison of quantum memory performances in the no-reset scheme using standard (triangles and dashed lines) and round-squeezing (diamonds and dotted lines) circuits. For each distance $d = 5,7,9$ (corresponding to red, blue, black colours), we performed $d$ QEC rounds.
    }
    \label{fig:mem_threshold_swappy_gadget}
\end{figure}

\begin{figure}[h!]
    \centering
    \includegraphics[width=0.48\textwidth]{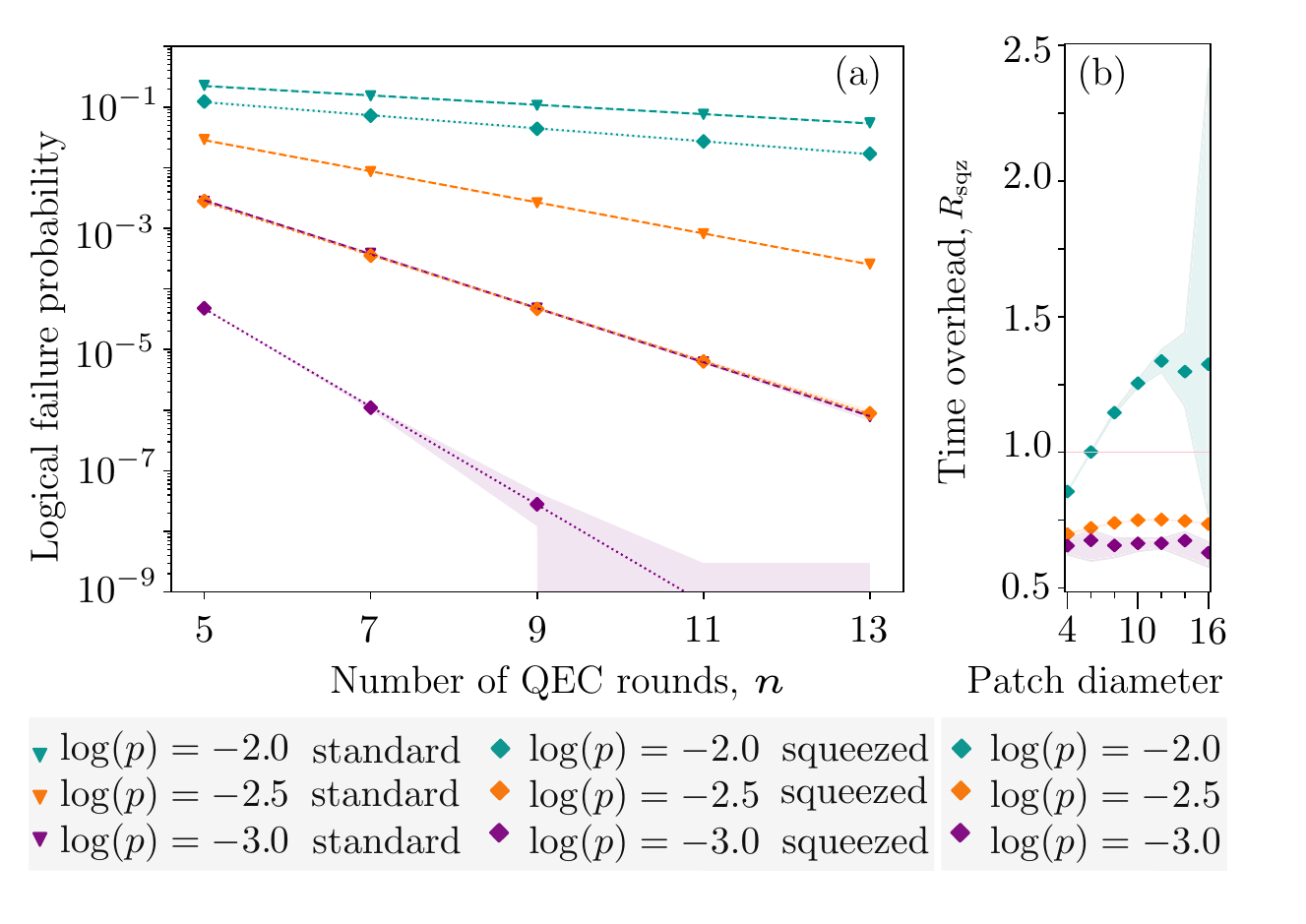}
    \caption{
    Comparison of stability experiment performances in the no-reset scheme using standard (triangles and dashed lines) and round-squeezing (diamonds and dotted lines) circuits. Colours indicate different physical error probabilities: $p = 10^{-2}$ (blue), $p = 10^{-2.5}$ (orange), $p = 10^{-3}$ (purple). 
    (a) Comparison of logical failure probabilities for a $4\times 4$ stability patch. Error bars as in \Cref{fig:stability_no_gadget_plots}(a) are also shown as shaded areas. Note that the standard $p = 10^{-3}$ and round-squeezing $p = 10^{-2.5}$ plots overlap. 
    (b) The time overhead $R_{\mathrm{sqz}}$ that the round-squeezing circuit requires when compared to the no-reset standard circuit. The shaded areas correspond to $90\%$ confidence intervals.}
    \label{fig:stability_swappy_gadget_plots}
\end{figure}

We simulated these circuits in order to compare the performance to the previously discussed syndrome extraction circuits. We applied the same principles for the number of shots as in \Cref{sec:qec_experiments,sec:spread}. We present the quantum memory results in \Cref{fig:mem_threshold_swappy_gadget} and see that again the performance is slightly diminished from the standard no-reset case, though not as much as with the error-spreading circuits. We attribute this to the increased number of two-qubit gates used in these syndrome extraction circuits. However, as we see from the stability time overhead plots in \Cref{fig:stability_swappy_gadget_plots}(b), we need fewer QEC rounds for the round-squeezing circuit to avoid undetectable time-like logical failures and so we need fewer QEC rounds for quantum memory as well, making the gap in \Cref{fig:mem_threshold_swappy_gadget} smaller. For instance, with $p=0.001$ we have $R_{\mathrm{sqz}}\approx 0.65$. Since execution times per QEC round differ, we need only $\approx 55\%$ of the number of QEC rounds with the round-squeezing circuit -- slightly more than half as many as for the standard no-reset case. Based on this, we can conclude that this circuit improves on the QEC performance of both the standard and error-spreading no-reset circuits, though with the use of additional physical qubits.

From comparing \Cref{fig:stability_swappy_gadget_plots}(b) and \Cref{fig:stability_no_gadget_plots}(b), we can further conclude that this circuit is a good alternative to implementing unconditional resets, even with instantaneous reset, at least for the range of physical error parameters we consider. We expect that, with much lower physical error probability, this would no longer be the case.

Whilst no device with the connectivity shown in \Cref{fig:swappy_gadget}(a) currently exists, it has lower connectivity than many current devices suggesting it should be easier to build. However, we point out that using an additional SWAP layer on one stabiliser type, it is possible to align the auxiliary qubits in a parallel direction without introducing ``bad hooks'' \cite{Wootton-Hetenyi}. In this case, the required connectivity is a sub-graph of the popular square-grid connectivity.

\section{Conclusion}\label{sec:conclusion}

We investigated the QEC consequences of performing mid-circuit reset, or not, during QEC experiments using standard syndrome extraction circuits.  Our main conclusion is that unconditionally resetting qubits, for instance by driving interactions with a dissipative environment, fundamentally improves the fault-tolerance properties of logical operations offering up-to $2\times$ speedup of fault-tolerant quantum computers. Access to fast, high-fidelity unconditional resets would make this approach a clear winner.

Nevertheless, with current best-in-class devices operating close to the QEC threshold and focused more on memory experiments, our simulations indicated that not resetting qubits can result in better performance. This justifies the approach taken in recent proof-of-principle experiments \cite{Delft-13, ETH-planar-code} and offers encouragement for near-term experiments to continue this practice. For those keen to avoid developing unconditional-reset technology, we presented two alternative syndrome extraction circuits that provide additional protection against measurement errors. These alternative circuits are interesting near-term experiments and potential long-term solutions. Whether our alternative circuits are genuine contenders for the best approach to FTQC will depend on additional effects, such as leakage errors, not considered in this paper. Such investigation would be interesting future work.

Furthermore, we emphasise again that our results are based on a particular noise model which assumes, for example, the measurement classification and qubit bit-flip error rates are 20$\%$ and 80$\%$ of the total measurement error rate, respectively. A direction for future work could be to repeat the analysis with different noise models or obtain results from real quantum computers. In particular, varying the measurement time on a particular device could affect results, not just by changing the amount of idling error on other qubits, but also by altering the total measurement error and the above split into classification and qubit bit-flip error.

As for the third class of reset strategy, conditional reset, our analysis shows that it will never compete with other strategies, and so should not be used in any of the QEC experiments considered here. This observation debunks folklore claims that control systems for QEC require fast conditional reset.

A summary of our results can be found in \Cref{tab:decision}, where we present the best-performing scheme in different regimes based on our circuit-level noise simulations. This table can be used to help decide which scheme to use for FTQC.

\begin{table}[]
\footnotesize
   \centering
    \begin{tabular}{ |c|c|c| } 
    \hline
    Physical error probability & Reset speed & Suggested scheme to use \\
    \hline
     & Fast & No reset \\ 
     \cline{2-3}
    High & Slow or & No reset \\ 
     &not available & \\
    \hline
     & & Unconditional reset,\\
     & Fast & error-spreading or \\
     Medium & & round-squeezing \\ 
    \cline{2-3}
    & Slow or & Error-spreading or \\ 
    & not available & round-squeezing \\
    \hline
     & Fast & Unconditional reset \\ 
     \cline{2-3}
    Low & Slow or & Error-spreading or \\
    & not available & round-squeezing \\ 
    \hline
    \end{tabular}
    \caption{Summary of the best-performing scheme(s) in different regimes based on our circuit-level noise simulations. This can be used to help decide which scheme to use for FTQC. For the `Physical error probability' column, `high' corresponds to $p\approx 10^{-2}$, `medium' to $p\approx 10^{-2.5}$ and `low' to $p\lessapprox 10^{-3}$. In the `Reset speed' column, `fast' corresponds to $\approx 100$ ns and `slow' to $\approx 500$ ns.}
    \label{tab:decision}
\end{table}

\bibliography{main}

\begin{thebibliography}{37}%
\makeatletter
\providecommand \@ifxundefined [1]{%
 \@ifx{#1\undefined}
}%
\providecommand \@ifnum [1]{%
 \ifnum #1\expandafter \@firstoftwo
 \else \expandafter \@secondoftwo
 \fi
}%
\providecommand \@ifx [1]{%
 \ifx #1\expandafter \@firstoftwo
 \else \expandafter \@secondoftwo
 \fi
}%
\providecommand \natexlab [1]{#1}%
\providecommand \enquote  [1]{``#1''}%
\providecommand \bibnamefont  [1]{#1}%
\providecommand \bibfnamefont [1]{#1}%
\providecommand \citenamefont [1]{#1}%
\providecommand \href@noop [0]{\@secondoftwo}%
\providecommand \href [0]{\begingroup \@sanitize@url \@href}%
\providecommand \@href[1]{\@@startlink{#1}\@@href}%
\providecommand \@@href[1]{\endgroup#1\@@endlink}%
\providecommand \@sanitize@url [0]{\catcode `\\12\catcode `\$12\catcode `\&12\catcode `\#12\catcode `\^12\catcode `\_12\catcode `\%12\relax}%
\providecommand \@@startlink[1]{}%
\providecommand \@@endlink[0]{}%
\providecommand \url  [0]{\begingroup\@sanitize@url \@url }%
\providecommand \@url [1]{\endgroup\@href {#1}{\urlprefix }}%
\providecommand \urlprefix  [0]{URL }%
\providecommand \Eprint [0]{\href }%
\providecommand \doibase [0]{https://doi.org/}%
\providecommand \selectlanguage [0]{\@gobble}%
\providecommand \bibinfo  [0]{\@secondoftwo}%
\providecommand \bibfield  [0]{\@secondoftwo}%
\providecommand \translation [1]{[#1]}%
\providecommand \BibitemOpen [0]{}%
\providecommand \bibitemStop [0]{}%
\providecommand \bibitemNoStop [0]{.\EOS\space}%
\providecommand \EOS [0]{\spacefactor3000\relax}%
\providecommand \BibitemShut  [1]{\csname bibitem#1\endcsname}%
\let\auto@bib@innerbib\@empty
\bibitem [{\citenamefont {Nielsen}\ and\ \citenamefont {Chuang}(2010)}]{NielsenChuang}%
  \BibitemOpen
  \bibfield  {author} {\bibinfo {author} {\bibfnamefont {M.~A.}\ \bibnamefont {Nielsen}}\ and\ \bibinfo {author} {\bibfnamefont {I.~L.}\ \bibnamefont {Chuang}},\ }\href@noop {} {\emph {\bibinfo {title} {Quantum Computation and Quantum Information: 10th Anniversary Edition}}}\ (\bibinfo  {publisher} {Cambridge University Press},\ \bibinfo {year} {2010})\BibitemShut {NoStop}%
\bibitem [{\citenamefont {DiVincenzo}(2000)}]{DiVincenzo}%
  \BibitemOpen
  \bibfield  {author} {\bibinfo {author} {\bibfnamefont {D.~P.}\ \bibnamefont {DiVincenzo}},\ }\bibfield  {title} {\bibinfo {title} {The physical implementation of quantum computation},\ }\href {https://doi.org/10.1002/1521-3978(200009)48:9/11<771::AID-PROP771>3.0.CO;2-E} {\bibfield  {journal} {\bibinfo  {journal} {Progr. Phys. Fortschr. Phys.}\ }\textbf {\bibinfo {volume} {48}},\ \bibinfo {pages} {771–783} (\bibinfo {year} {2000})}\BibitemShut {NoStop}%
\bibitem [{\citenamefont {Magnard}\ \emph {et~al.}(2018)\citenamefont {Magnard}, \citenamefont {Kurpiers}, \citenamefont {Royer} \emph {et~al.}}]{fast-and-unconditional-reset}%
  \BibitemOpen
  \bibfield  {author} {\bibinfo {author} {\bibfnamefont {P.}~\bibnamefont {Magnard}}, \bibinfo {author} {\bibfnamefont {P.}~\bibnamefont {Kurpiers}}, \bibinfo {author} {\bibfnamefont {B.}~\bibnamefont {Royer}}, \emph {et~al.},\ }\bibfield  {title} {\bibinfo {title} {Fast and unconditional all-microwave reset of a superconducting qubit},\ }\href {https://doi.org/10.1103/PhysRevLett.121.060502} {\bibfield  {journal} {\bibinfo  {journal} {Phys. Rev. Lett.}\ }\textbf {\bibinfo {volume} {121}},\ \bibinfo {pages} {060502} (\bibinfo {year} {2018})}\BibitemShut {NoStop}%
\bibitem [{\citenamefont {Zhou}\ \emph {et~al.}(2021)\citenamefont {Zhou}, \citenamefont {Zhang}, \citenamefont {Yin} \emph {et~al.}}]{rapid-unconditional-reset}%
  \BibitemOpen
  \bibfield  {author} {\bibinfo {author} {\bibfnamefont {Y.}~\bibnamefont {Zhou}}, \bibinfo {author} {\bibfnamefont {Z.}~\bibnamefont {Zhang}}, \bibinfo {author} {\bibfnamefont {Z.}~\bibnamefont {Yin}}, \emph {et~al.},\ }\bibfield  {title} {\bibinfo {title} {Rapid and unconditional parametric reset protocol for tunable superconducting qubits},\ }\href {https://doi.org/10.1038/s41467-021-26205-y} {\bibfield  {journal} {\bibinfo  {journal} {Nat Commun}\ }\textbf {\bibinfo {volume} {12}},\ \bibinfo {pages} {5924} (\bibinfo {year} {2021})}\BibitemShut {NoStop}%
\bibitem [{\citenamefont {C\'orcoles}\ \emph {et~al.}(2021)\citenamefont {C\'orcoles}, \citenamefont {Takita}, \citenamefont {Inoue} \emph {et~al.}}]{conditional-reset}%
  \BibitemOpen
  \bibfield  {author} {\bibinfo {author} {\bibfnamefont {A.~D.}\ \bibnamefont {C\'orcoles}}, \bibinfo {author} {\bibfnamefont {M.}~\bibnamefont {Takita}}, \bibinfo {author} {\bibfnamefont {K.}~\bibnamefont {Inoue}}, \emph {et~al.},\ }\bibfield  {title} {\bibinfo {title} {Exploiting dynamic quantum circuits in a quantum algorithm with superconducting qubits},\ }\href {https://doi.org/10.1103/PhysRevLett.127.100501} {\bibfield  {journal} {\bibinfo  {journal} {Phys. Rev. Lett.}\ }\textbf {\bibinfo {volume} {127}},\ \bibinfo {pages} {100501} (\bibinfo {year} {2021})}\BibitemShut {NoStop}%
\bibitem [{\citenamefont {Ali}\ \emph {et~al.}(2024)\citenamefont {Ali}, \citenamefont {Marques}, \citenamefont {Crawford} \emph {et~al.}}]{Delft-13}%
  \BibitemOpen
  \bibfield  {author} {\bibinfo {author} {\bibfnamefont {H.}~\bibnamefont {Ali}}, \bibinfo {author} {\bibfnamefont {J.}~\bibnamefont {Marques}}, \bibinfo {author} {\bibfnamefont {O.}~\bibnamefont {Crawford}}, \emph {et~al.},\ }\bibfield  {title} {\bibinfo {title} {Reducing the error rate of a superconducting logical qubit using analog readout information},\ }\href {https://doi.org/10.1103/PhysRevApplied.22.044031} {\bibfield  {journal} {\bibinfo  {journal} {Phys. Rev. Appl.}\ }\textbf {\bibinfo {volume} {22}},\ \bibinfo {pages} {044031} (\bibinfo {year} {2024})}\BibitemShut {NoStop}%
\bibitem [{\citenamefont {Marques}\ \emph {et~al.}(2022)\citenamefont {Marques}, \citenamefont {Varbanov}, \citenamefont {Moreira} \emph {et~al.}}]{Delft-planar-docde}%
  \BibitemOpen
  \bibfield  {author} {\bibinfo {author} {\bibfnamefont {J.~F.}\ \bibnamefont {Marques}}, \bibinfo {author} {\bibfnamefont {B.~M.}\ \bibnamefont {Varbanov}}, \bibinfo {author} {\bibfnamefont {M.~S.}\ \bibnamefont {Moreira}}, \emph {et~al.},\ }\bibfield  {title} {\bibinfo {title} {Logical-qubit operations in an error-detecting surface code},\ }\href {https://doi.org/https://doi.org/10.1038/s41567-021-01423-9} {\bibfield  {journal} {\bibinfo  {journal} {Nature Physics}\ }\textbf {\bibinfo {volume} {18}},\ \bibinfo {pages} {80–86} (\bibinfo {year} {2022})}\BibitemShut {NoStop}%
\bibitem [{\citenamefont {Krinner}\ \emph {et~al.}(2022)\citenamefont {Krinner}, \citenamefont {Lacroix}, \citenamefont {Remm} \emph {et~al.}}]{ETH-planar-code}%
  \BibitemOpen
  \bibfield  {author} {\bibinfo {author} {\bibfnamefont {S.}~\bibnamefont {Krinner}}, \bibinfo {author} {\bibfnamefont {N.}~\bibnamefont {Lacroix}}, \bibinfo {author} {\bibfnamefont {A.}~\bibnamefont {Remm}}, \emph {et~al.},\ }\bibfield  {title} {\bibinfo {title} {Realizing repeated quantum error correction in a distance-three surface code},\ }\href {https://doi.org/https://doi.org/10.1038/s41586-022-04566-8} {\bibfield  {journal} {\bibinfo  {journal} {Nature}\ }\textbf {\bibinfo {volume} {605}},\ \bibinfo {pages} {669–674} (\bibinfo {year} {2022})}\BibitemShut {NoStop}%
\bibitem [{\citenamefont {Campbell}(2024)}]{campbell2024series}%
  \BibitemOpen
  \bibfield  {author} {\bibinfo {author} {\bibfnamefont {E.}~\bibnamefont {Campbell}},\ }\bibfield  {title} {\bibinfo {title} {A series of fast-paced advances in quantum error correction},\ }\href {https://doi.org/10.1038/s42254-024-00706-3} {\bibfield  {journal} {\bibinfo  {journal} {Nature Reviews Physics}\ }\textbf {\bibinfo {volume} {6}},\ \bibinfo {pages} {160} (\bibinfo {year} {2024})}\BibitemShut {NoStop}%
\bibitem [{\citenamefont {Bluvstein}\ \emph {et~al.}(2024)\citenamefont {Bluvstein}, \citenamefont {Evered}, \citenamefont {Geim} \emph {et~al.}}]{quera-paper}%
  \BibitemOpen
  \bibfield  {author} {\bibinfo {author} {\bibfnamefont {D.}~\bibnamefont {Bluvstein}}, \bibinfo {author} {\bibfnamefont {S.}~\bibnamefont {Evered}}, \bibinfo {author} {\bibfnamefont {A.}~\bibnamefont {Geim}}, \emph {et~al.},\ }\bibfield  {title} {\bibinfo {title} {Logical quantum processor based on reconfigurable atom arrays},\ }\href {https://doi.org/10.1038/s41586-023-06927-3} {\bibfield  {journal} {\bibinfo  {journal} {Nature}\ }\textbf {\bibinfo {volume} {626}},\ \bibinfo {pages} {58–65} (\bibinfo {year} {2024})}\BibitemShut {NoStop}%
\bibitem [{\citenamefont {Ryan-Anderson}\ \emph {et~al.}(2022)\citenamefont {Ryan-Anderson}, \citenamefont {Brown}, \citenamefont {Allman} \emph {et~al.}}]{quantinuum-paper}%
  \BibitemOpen
  \bibfield  {author} {\bibinfo {author} {\bibfnamefont {C.}~\bibnamefont {Ryan-Anderson}}, \bibinfo {author} {\bibfnamefont {N.~C.}\ \bibnamefont {Brown}}, \bibinfo {author} {\bibfnamefont {M.~S.}\ \bibnamefont {Allman}}, \emph {et~al.},\ }\bibfield  {title} {\bibinfo {title} {Implementing fault-tolerant entangling gates on the five-qubit code and the color code},\ }\href {https://arxiv.org/abs/2208.01863} {\bibfield  {journal} {\bibinfo  {journal} {arXiv}\ } (\bibinfo {year} {2022})}\BibitemShut {NoStop}%
\bibitem [{\citenamefont {Menendez}\ \emph {et~al.}(2023)\citenamefont {Menendez}, \citenamefont {Ray},\ and\ \citenamefont {Vasmer}}]{832-color-code-demonstration}%
  \BibitemOpen
  \bibfield  {author} {\bibinfo {author} {\bibfnamefont {D.}~\bibnamefont {Menendez}}, \bibinfo {author} {\bibfnamefont {A.}~\bibnamefont {Ray}},\ and\ \bibinfo {author} {\bibfnamefont {M.}~\bibnamefont {Vasmer}},\ }\bibfield  {title} {\bibinfo {title} {Implementing fault-tolerant non-clifford gates using the [[8,3,2]] color code},\ }\href {https://arxiv.org/abs/2309.08663} {\bibfield  {journal} {\bibinfo  {journal} {arXiv}\ } (\bibinfo {year} {2023})}\BibitemShut {NoStop}%
\bibitem [{\citenamefont {Miao}\ \emph {et~al.}(2023)\citenamefont {Miao}, \citenamefont {McEwen}, \citenamefont {Atalaya} \emph {et~al.}}]{miao_overcoming_2022}%
  \BibitemOpen
  \bibfield  {author} {\bibinfo {author} {\bibfnamefont {K.~C.}\ \bibnamefont {Miao}}, \bibinfo {author} {\bibfnamefont {M.}~\bibnamefont {McEwen}}, \bibinfo {author} {\bibfnamefont {J.}~\bibnamefont {Atalaya}}, \emph {et~al.},\ }\bibfield  {title} {\bibinfo {title} {Overcoming leakage in scalable quantum error correction},\ }\href {https://doi.org/https://doi.org/10.1038/s41567-023-02226-w} {\bibfield  {journal} {\bibinfo  {journal} {Nature Physics}\ }\textbf {\bibinfo {volume} {19}},\ \bibinfo {pages} {1780–1786} (\bibinfo {year} {2023})}\BibitemShut {NoStop}%
\bibitem [{\citenamefont {Acharya}\ \emph {et~al.}(2023)\citenamefont {Acharya}, \citenamefont {Aleiner}, \citenamefont {Allen} \emph {et~al.}}]{acharya_suppressing_2023}%
  \BibitemOpen
  \bibfield  {author} {\bibinfo {author} {\bibfnamefont {R.}~\bibnamefont {Acharya}}, \bibinfo {author} {\bibfnamefont {I.}~\bibnamefont {Aleiner}}, \bibinfo {author} {\bibfnamefont {R.}~\bibnamefont {Allen}}, \emph {et~al.},\ }\bibfield  {title} {\bibinfo {title} {Suppressing quantum errors by scaling a surface code logical qubit},\ }\href {https://doi.org/10.1038/s41586-022-05434-1} {\bibfield  {journal} {\bibinfo  {journal} {Nature}\ }\textbf {\bibinfo {volume} {614}},\ \bibinfo {pages} {676} (\bibinfo {year} {2023})}\BibitemShut {NoStop}%
\bibitem [{\citenamefont {Horsman}\ \emph {et~al.}(2012)\citenamefont {Horsman}, \citenamefont {Fowler}, \citenamefont {Devitt},\ and\ \citenamefont {Van~Meter}}]{Horsman_2012}%
  \BibitemOpen
  \bibfield  {author} {\bibinfo {author} {\bibfnamefont {D.}~\bibnamefont {Horsman}}, \bibinfo {author} {\bibfnamefont {A.~G.}\ \bibnamefont {Fowler}}, \bibinfo {author} {\bibfnamefont {S.}~\bibnamefont {Devitt}},\ and\ \bibinfo {author} {\bibfnamefont {R.}~\bibnamefont {Van~Meter}},\ }\bibfield  {title} {\bibinfo {title} {Surface code quantum computing by lattice surgery},\ }\href {https://doi.org/10.1088/1367-2630/14/12/123011} {\bibfield  {journal} {\bibinfo  {journal} {New Journal of Physics}\ }\textbf {\bibinfo {volume} {14}},\ \bibinfo {pages} {123011} (\bibinfo {year} {2012})}\BibitemShut {NoStop}%
\bibitem [{\citenamefont {Litinski}(2019)}]{GoSc}%
  \BibitemOpen
  \bibfield  {author} {\bibinfo {author} {\bibfnamefont {D.}~\bibnamefont {Litinski}},\ }\bibfield  {title} {\bibinfo {title} {A game of surface codes: Large-scale quantum computing with lattice surgery},\ }\href {https://doi.org/10.22331/q-2019-03-05-128} {\bibfield  {journal} {\bibinfo  {journal} {Quantum}\ }\textbf {\bibinfo {volume} {3}},\ \bibinfo {pages} {128} (\bibinfo {year} {2019})}\BibitemShut {NoStop}%
\bibitem [{\citenamefont {Chamberland}\ and\ \citenamefont {Campbell}(2022{\natexlab{a}})}]{ChamCamtwistfree}%
  \BibitemOpen
  \bibfield  {author} {\bibinfo {author} {\bibfnamefont {C.}~\bibnamefont {Chamberland}}\ and\ \bibinfo {author} {\bibfnamefont {E.~T.}\ \bibnamefont {Campbell}},\ }\bibfield  {title} {\bibinfo {title} {Universal quantum computing with twist-free and temporally encoded lattice surgery},\ }\href {https://doi.org/10.1103/PRXQuantum.3.010331} {\bibfield  {journal} {\bibinfo  {journal} {PRX Quantum}\ }\textbf {\bibinfo {volume} {3}},\ \bibinfo {pages} {010331} (\bibinfo {year} {2022}{\natexlab{a}})}\BibitemShut {NoStop}%
\bibitem [{\citenamefont {Chamberland}\ and\ \citenamefont {Campbell}(2022{\natexlab{b}})}]{ChamCamtwistbased}%
  \BibitemOpen
  \bibfield  {author} {\bibinfo {author} {\bibfnamefont {C.}~\bibnamefont {Chamberland}}\ and\ \bibinfo {author} {\bibfnamefont {E.~T.}\ \bibnamefont {Campbell}},\ }\bibfield  {title} {\bibinfo {title} {Circuit-level protocol and analysis for twist-based lattice surgery},\ }\href {https://doi.org/10.1103/PhysRevResearch.4.023090} {\bibfield  {journal} {\bibinfo  {journal} {Physical Review Research}\ }\textbf {\bibinfo {volume} {4}},\ \bibinfo {pages} {023090} (\bibinfo {year} {2022}{\natexlab{b}})}\BibitemShut {NoStop}%
\bibitem [{\citenamefont {Bombin}\ \emph {et~al.}(2023)\citenamefont {Bombin}, \citenamefont {Dawson}, \citenamefont {Mishmash} \emph {et~al.}}]{Bombin2021}%
  \BibitemOpen
  \bibfield  {author} {\bibinfo {author} {\bibfnamefont {H.}~\bibnamefont {Bombin}}, \bibinfo {author} {\bibfnamefont {C.}~\bibnamefont {Dawson}}, \bibinfo {author} {\bibfnamefont {R.~V.}\ \bibnamefont {Mishmash}}, \emph {et~al.},\ }\bibfield  {title} {\bibinfo {title} {Logical blocks for fault-tolerant topological quantum computation},\ }\href {https://doi.org/10.1103/PRXQuantum.4.020303} {\bibfield  {journal} {\bibinfo  {journal} {PRX Quantum}\ }\textbf {\bibinfo {volume} {4}},\ \bibinfo {pages} {020303} (\bibinfo {year} {2023})}\BibitemShut {NoStop}%
\bibitem [{\citenamefont {Geh\'er}\ \emph {et~al.}(2024{\natexlab{a}})\citenamefont {Geh\'er}, \citenamefont {McLauchlan}, \citenamefont {Campbell} \emph {et~al.}}]{geher_error-corrected_2023}%
  \BibitemOpen
  \bibfield  {author} {\bibinfo {author} {\bibfnamefont {G.~P.}\ \bibnamefont {Geh\'er}}, \bibinfo {author} {\bibfnamefont {C.}~\bibnamefont {McLauchlan}}, \bibinfo {author} {\bibfnamefont {E.~T.}\ \bibnamefont {Campbell}}, \emph {et~al.},\ }\bibfield  {title} {\bibinfo {title} {Error-corrected {Hadamard} gate simulated at the circuit level},\ }\href {https://doi.org/https://doi.org/10.22331/q-2024-07-02-1394} {\bibfield  {journal} {\bibinfo  {journal} {Quantum}\ }\textbf {\bibinfo {volume} {8}},\ \bibinfo {pages} {1394} (\bibinfo {year} {2024}{\natexlab{a}})}\BibitemShut {NoStop}%
\bibitem [{\citenamefont {Gidney}(2022)}]{Gidney-stability}%
  \BibitemOpen
  \bibfield  {author} {\bibinfo {author} {\bibfnamefont {C.}~\bibnamefont {Gidney}},\ }\bibfield  {title} {\bibinfo {title} {Stability {E}xperiments: {T}he {O}verlooked {D}ual of {M}emory {E}xperiments},\ }\href {https://doi.org/10.22331/q-2022-08-24-786} {\bibfield  {journal} {\bibinfo  {journal} {{Quantum}}\ }\textbf {\bibinfo {volume} {6}},\ \bibinfo {pages} {786} (\bibinfo {year} {2022})}\BibitemShut {NoStop}%
\bibitem [{\citenamefont {Chao}\ and\ \citenamefont {Reichardt}(2020)}]{flag}%
  \BibitemOpen
  \bibfield  {author} {\bibinfo {author} {\bibfnamefont {R.}~\bibnamefont {Chao}}\ and\ \bibinfo {author} {\bibfnamefont {B.~W.}\ \bibnamefont {Reichardt}},\ }\bibfield  {title} {\bibinfo {title} {Flag fault-tolerant error correction for any stabilizer code},\ }\href {https://doi.org/10.1103/PRXQuantum.1.010302} {\bibfield  {journal} {\bibinfo  {journal} {PRX Quantum}\ }\textbf {\bibinfo {volume} {1}},\ \bibinfo {pages} {010302} (\bibinfo {year} {2020})}\BibitemShut {NoStop}%
\bibitem [{\citenamefont {McEwen}\ \emph {et~al.}(2023)\citenamefont {McEwen}, \citenamefont {Bacon},\ and\ \citenamefont {Gidney}}]{McEwenBaconGidney}%
  \BibitemOpen
  \bibfield  {author} {\bibinfo {author} {\bibfnamefont {M.}~\bibnamefont {McEwen}}, \bibinfo {author} {\bibfnamefont {D.}~\bibnamefont {Bacon}},\ and\ \bibinfo {author} {\bibfnamefont {C.}~\bibnamefont {Gidney}},\ }\bibfield  {title} {\bibinfo {title} {Relaxing hardware requirements for surface code circuits using time-dynamics},\ }\href {https://doi.org/https://doi.org/10.22331/q-2023-11-07-1172} {\bibfield  {journal} {\bibinfo  {journal} {Quantum}\ }\textbf {\bibinfo {volume} {7}},\ \bibinfo {pages} {1172} (\bibinfo {year} {2023})}\BibitemShut {NoStop}%
\bibitem [{\citenamefont {Dennis}\ \emph {et~al.}(2002)\citenamefont {Dennis}, \citenamefont {Kitaev}, \citenamefont {Landahl},\ and\ \citenamefont {Preskill}}]{Dennis-Kitaev-Landahl}%
  \BibitemOpen
  \bibfield  {author} {\bibinfo {author} {\bibfnamefont {E.}~\bibnamefont {Dennis}}, \bibinfo {author} {\bibfnamefont {A.}~\bibnamefont {Kitaev}}, \bibinfo {author} {\bibfnamefont {A.}~\bibnamefont {Landahl}},\ and\ \bibinfo {author} {\bibfnamefont {J.}~\bibnamefont {Preskill}},\ }\bibfield  {title} {\bibinfo {title} {{Topological quantum memory}},\ }\href {https://doi.org/10.1063/1.1499754} {\bibfield  {journal} {\bibinfo  {journal} {Journal of Mathematical Physics}\ }\textbf {\bibinfo {volume} {43}},\ \bibinfo {pages} {4452} (\bibinfo {year} {2002})}\BibitemShut {NoStop}%
\bibitem [{\citenamefont {Higgott}\ and\ \citenamefont {Gidney}(2023)}]{pymatching}%
  \BibitemOpen
  \bibfield  {author} {\bibinfo {author} {\bibfnamefont {O.}~\bibnamefont {Higgott}}\ and\ \bibinfo {author} {\bibfnamefont {C.}~\bibnamefont {Gidney}},\ }\bibfield  {title} {\bibinfo {title} {Sparse blossom: correcting a million errors per core second with minimum-weight matching},\ }\bibfield  {journal} {\bibinfo  {journal} {{arXiv:2303.15933}}\ }\href {https://doi.org/https://doi.org/10.48550/arXiv.2303.15933} {https://doi.org/10.48550/arXiv.2303.15933} (\bibinfo {year} {2023})\BibitemShut {NoStop}%
\bibitem [{\citenamefont {Varbanov}\ \emph {et~al.}(2020)\citenamefont {Varbanov}, \citenamefont {Battistel}, \citenamefont {Tarasinski} \emph {et~al.}}]{varbanov2020leakage}%
  \BibitemOpen
  \bibfield  {author} {\bibinfo {author} {\bibfnamefont {B.~M.}\ \bibnamefont {Varbanov}}, \bibinfo {author} {\bibfnamefont {F.}~\bibnamefont {Battistel}}, \bibinfo {author} {\bibfnamefont {B.~M.}\ \bibnamefont {Tarasinski}}, \emph {et~al.},\ }\bibfield  {title} {\bibinfo {title} {Leakage detection for a transmon-based surface code},\ }\href@noop {} {\bibfield  {journal} {\bibinfo  {journal} {npj Quantum Information}\ }\textbf {\bibinfo {volume} {6}},\ \bibinfo {pages} {102} (\bibinfo {year} {2020})}\BibitemShut {NoStop}%
\bibitem [{\citenamefont {Sahay}\ \emph {et~al.}(2024)\citenamefont {Sahay}, \citenamefont {Lin}, \citenamefont {Huang}, \citenamefont {Brown},\ and\ \citenamefont {Puri}}]{sahay2024error}%
  \BibitemOpen
  \bibfield  {author} {\bibinfo {author} {\bibfnamefont {K.}~\bibnamefont {Sahay}}, \bibinfo {author} {\bibfnamefont {Y.}~\bibnamefont {Lin}}, \bibinfo {author} {\bibfnamefont {S.}~\bibnamefont {Huang}}, \bibinfo {author} {\bibfnamefont {K.~R.}\ \bibnamefont {Brown}},\ and\ \bibinfo {author} {\bibfnamefont {S.}~\bibnamefont {Puri}},\ }\bibfield  {title} {\bibinfo {title} {Error correction of transversal cnot gates for scalable surface code computation},\ }\href@noop {} {\bibfield  {journal} {\bibinfo  {journal} {arXiv preprint arXiv:2408.01393}\ } (\bibinfo {year} {2024})}\BibitemShut {NoStop}%
\bibitem [{\citenamefont {Wan}\ \emph {et~al.}(2024)\citenamefont {Wan}, \citenamefont {Webber}, \citenamefont {Fowler},\ and\ \citenamefont {Hensinger}}]{wan2024iterative}%
  \BibitemOpen
  \bibfield  {author} {\bibinfo {author} {\bibfnamefont {K.~H.}\ \bibnamefont {Wan}}, \bibinfo {author} {\bibfnamefont {M.}~\bibnamefont {Webber}}, \bibinfo {author} {\bibfnamefont {A.~G.}\ \bibnamefont {Fowler}},\ and\ \bibinfo {author} {\bibfnamefont {W.~K.}\ \bibnamefont {Hensinger}},\ }\bibfield  {title} {\bibinfo {title} {An iterative transversal cnot decoder},\ }\href@noop {} {\bibfield  {journal} {\bibinfo  {journal} {arXiv preprint arXiv:2407.20976}\ } (\bibinfo {year} {2024})}\BibitemShut {NoStop}%
\bibitem [{\citenamefont {Zhou}\ \emph {et~al.}(2024)\citenamefont {Zhou}, \citenamefont {Zhao}, \citenamefont {Cain}, \citenamefont {Bluvstein}, \citenamefont {Duckering}, \citenamefont {Hu}, \citenamefont {Wang}, \citenamefont {Kubica},\ and\ \citenamefont {Lukin}}]{zhou2024algorithmic}%
  \BibitemOpen
  \bibfield  {author} {\bibinfo {author} {\bibfnamefont {H.}~\bibnamefont {Zhou}}, \bibinfo {author} {\bibfnamefont {C.}~\bibnamefont {Zhao}}, \bibinfo {author} {\bibfnamefont {M.}~\bibnamefont {Cain}}, \bibinfo {author} {\bibfnamefont {D.}~\bibnamefont {Bluvstein}}, \bibinfo {author} {\bibfnamefont {C.}~\bibnamefont {Duckering}}, \bibinfo {author} {\bibfnamefont {H.-Y.}\ \bibnamefont {Hu}}, \bibinfo {author} {\bibfnamefont {S.-T.}\ \bibnamefont {Wang}}, \bibinfo {author} {\bibfnamefont {A.}~\bibnamefont {Kubica}},\ and\ \bibinfo {author} {\bibfnamefont {M.~D.}\ \bibnamefont {Lukin}},\ }\bibfield  {title} {\bibinfo {title} {Algorithmic fault tolerance for fast quantum computing},\ }\href@noop {} {\bibfield  {journal} {\bibinfo  {journal} {arXiv preprint arXiv:2406.17653}\ } (\bibinfo {year} {2024})}\BibitemShut {NoStop}%
\bibitem [{Rig(2023)}]{Rigetti-Ankaa}%
  \BibitemOpen
  \href@noop {} {\bibinfo {title} {{Rigetti Systems}}},\ \bibinfo {howpublished} {\url{https://qcs.rigetti.com/qpus}} (\bibinfo {year} {2023}),\ \bibinfo {note} {accessed: 01/07/2023}\BibitemShut {NoStop}%
\bibitem [{\citenamefont {Geh\'er}\ \emph {et~al.}(2024{\natexlab{b}})\citenamefont {Geh\'er}, \citenamefont {Jastrzebski}, \citenamefont {Campbell},\ and\ \citenamefont {Crawford}}]{our_stim_circuits}%
  \BibitemOpen
  \bibfield  {author} {\bibinfo {author} {\bibfnamefont {G.~P.}\ \bibnamefont {Geh\'er}}, \bibinfo {author} {\bibfnamefont {M.}~\bibnamefont {Jastrzebski}}, \bibinfo {author} {\bibfnamefont {E.~T.}\ \bibnamefont {Campbell}},\ and\ \bibinfo {author} {\bibfnamefont {O.}~\bibnamefont {Crawford}},\ }\href {https://doi.org/https://doi.org/10.5281/zenodo.13152440} {\bibinfo {title} {{Stim circuits for ``To reset, or not to reset -- that is the question" manuscript}}} (\bibinfo {year} {2024}{\natexlab{b}})\BibitemShut {NoStop}%
\bibitem [{\citenamefont {Gidney}\ \emph {et~al.}(2021)\citenamefont {Gidney}, \citenamefont {Newman}, \citenamefont {Fowler},\ and\ \citenamefont {Broughton}}]{gidney2021fault}%
  \BibitemOpen
  \bibfield  {author} {\bibinfo {author} {\bibfnamefont {C.}~\bibnamefont {Gidney}}, \bibinfo {author} {\bibfnamefont {M.}~\bibnamefont {Newman}}, \bibinfo {author} {\bibfnamefont {A.}~\bibnamefont {Fowler}},\ and\ \bibinfo {author} {\bibfnamefont {M.}~\bibnamefont {Broughton}},\ }\bibfield  {title} {\bibinfo {title} {A fault-tolerant honeycomb memory},\ }\href {https://doi.org/10.22331/q-2021-12-20-605} {\bibfield  {journal} {\bibinfo  {journal} {Quantum}\ }\textbf {\bibinfo {volume} {5}},\ \bibinfo {pages} {605} (\bibinfo {year} {2021})}\BibitemShut {NoStop}%
\bibitem [{\citenamefont {Sarvepalli}\ \emph {et~al.}(2009)\citenamefont {Sarvepalli}, \citenamefont {Klappenecker},\ and\ \citenamefont {R{\"o}tteler}}]{sarvepalli2009asymmetric}%
  \BibitemOpen
  \bibfield  {author} {\bibinfo {author} {\bibfnamefont {P.~K.}\ \bibnamefont {Sarvepalli}}, \bibinfo {author} {\bibfnamefont {A.}~\bibnamefont {Klappenecker}},\ and\ \bibinfo {author} {\bibfnamefont {M.}~\bibnamefont {R{\"o}tteler}},\ }\bibfield  {title} {\bibinfo {title} {Asymmetric quantum codes: constructions, bounds and performance},\ }\href@noop {} {\bibfield  {journal} {\bibinfo  {journal} {Proceedings of the Royal Society A: Mathematical, Physical and Engineering Sciences}\ }\textbf {\bibinfo {volume} {465}},\ \bibinfo {pages} {1645} (\bibinfo {year} {2009})}\BibitemShut {NoStop}%
\bibitem [{\citenamefont {Gidney}(2021)}]{stim}%
  \BibitemOpen
  \bibfield  {author} {\bibinfo {author} {\bibfnamefont {C.}~\bibnamefont {Gidney}},\ }\bibfield  {title} {\bibinfo {title} {Stim: a fast stabilizer circuit simulator},\ }\href {https://doi.org/10.22331/q-2021-07-06-497} {\bibfield  {journal} {\bibinfo  {journal} {{Quantum}}\ }\textbf {\bibinfo {volume} {5}},\ \bibinfo {pages} {497} (\bibinfo {year} {2021})}\BibitemShut {NoStop}%
\bibitem [{\citenamefont {Geh\'er}\ \emph {et~al.}(2024{\natexlab{c}})\citenamefont {Geh\'er}, \citenamefont {Crawford},\ and\ \citenamefont {Campbell}}]{geher_tangling_2023}%
  \BibitemOpen
  \bibfield  {author} {\bibinfo {author} {\bibfnamefont {G.~P.}\ \bibnamefont {Geh\'er}}, \bibinfo {author} {\bibfnamefont {O.}~\bibnamefont {Crawford}},\ and\ \bibinfo {author} {\bibfnamefont {E.~T.}\ \bibnamefont {Campbell}},\ }\bibfield  {title} {\bibinfo {title} {Tangling schedules eases hardware connectivity requirements for quantum error correction},\ }\href {https://doi.org/https://doi.org/10.1103/PRXQuantum.5.010348} {\bibfield  {journal} {\bibinfo  {journal} {PRX Quantum}\ }\textbf {\bibinfo {volume} {5}},\ \bibinfo {pages} {010348} (\bibinfo {year} {2024}{\natexlab{c}})}\BibitemShut {NoStop}%
\bibitem [{\citenamefont {Gidney}(2023)}]{Gidney-pentagonal}%
  \BibitemOpen
  \bibfield  {author} {\bibinfo {author} {\bibfnamefont {C.}~\bibnamefont {Gidney}},\ }\bibfield  {title} {\bibinfo {title} {A pair measurement surface code on pentagons},\ }\href {https://doi.org/https://doi.org/10.22331/q-2023-10-25-1156} {\bibfield  {journal} {\bibinfo  {journal} {Quantum}\ }\textbf {\bibinfo {volume} {7}},\ \bibinfo {pages} {1156} (\bibinfo {year} {2023})}\BibitemShut {NoStop}%
\bibitem [{\citenamefont {Het\'enyi}\ and\ \citenamefont {Wootton}(2024)}]{Wootton-Hetenyi}%
  \BibitemOpen
  \bibfield  {author} {\bibinfo {author} {\bibfnamefont {B.}~\bibnamefont {Het\'enyi}}\ and\ \bibinfo {author} {\bibfnamefont {J.~R.}\ \bibnamefont {Wootton}},\ }\bibfield  {title} {\bibinfo {title} {Tailoring quantum error correction to spin qubits},\ }\href {https://doi.org/10.1103/PhysRevA.109.032433} {\bibfield  {journal} {\bibinfo  {journal} {Phys. Rev. A}\ }\textbf {\bibinfo {volume} {109}},\ \bibinfo {pages} {032433} (\bibinfo {year} {2024})}\BibitemShut {NoStop}%
\end{thebibliography}%

\end{document}